\documentclass{article}

\usepackage[preprint]{neurips_2025}

\usepackage{xspace}
\usepackage{enumitem}
\usepackage{svg}
\usepackage{algorithm}
\usepackage{algpseudocode}
\usepackage{multirow}
\usepackage{subcaption}
\usepackage{flushend,balance}
\usepackage[simplified]{pgf-umlcd}
\usepackage{tikz}
\usepackage{pgfplots}
\usetikzlibrary{patterns}
\usepackage{color,tcolorbox,soul}
\usepackage{makecell}
\usepackage{arydshln}  %
\usepackage{bbm}
\usepackage{amsmath} 
\usepackage{wrapfig}

\newtheorem{theorem}{Theorem} %

\usepgfplotslibrary{groupplots}

\definecolor{americanrose}{rgb}{1.0, 0.70, 0.70}
\definecolor{airforceblue}{rgb}{0.70, 0.70, 1.00}
\definecolor{mygreen}{rgb}{0.40,0.761,0.651}
\definecolor{myorange}{rgb}{0.988,0.557,0.384}
\definecolor{mylightgreen}{rgb}{0.651,0.847,0.325}
\definecolor{myblue}{rgb}{0.557,0.627,0.796}
\definecolor{mypurple}{rgb}{0.906,0.541,0.765}
\definecolor{navyblue}{RGB}{0,0,128}

\newcommand{\system}{{\sc Needle}\xspace}
\newcommand{\cli}{{\tt needlectl}\xspace}

\newcommand{\lvis}{{\sc LVIS}\xspace}

\newcommand{\clip}{{\sc CLIP}\xspace}

\newcommand{\stitle}[1]{\noindent{\bf #1:}}
\newcommand{\at}[1]{{\tt \small #1}\xspace}

\newcommand{\dalle}{{\sc DALL·E 3}\xspace}

\newcommand{\gee}{\mathbf{g}}
\newcommand{\dee}{\mathcal{D}}
\newcommand{\dist}{\xi}

\newcommand{\ee}{\mathbb{E}}
\newcommand{\eps}{\varepsilon}
\newcommand{\embedding}{\mathcal{E}}
\newcommand{\nlq}{\varphi}

\renewcommand{\Re}{\mathbb{R}}%

\newcommand{\submitRM}[1]{}

\newcommand{\mnote}[1]{}  %
\newcommand{\innote}[1]{} 

\newcommand{\hard}[1]{\textcolor{navyblue}{#1}}

\newtcolorbox{exbox}{
  colback=blue!10,
  colframe=blue!20,
  arc=2mm,
  fonttitle=\bfseries,
  boxrule=0mm,
  boxsep=1mm,
  left=0mm,
  right=0mm,
  top=0mm,
  bottom=0mm
}

\usepackage[utf8]{inputenc} %
\usepackage[T1]{fontenc}    %
\usepackage{hyperref}       %
\usepackage{url}            %
\usepackage{booktabs}       %
\usepackage{amsfonts}       %
\usepackage{nicefrac}       %
\usepackage{microtype}      %
\usepackage{xcolor}         %

\title{\system: A Generative AI-Powered Multi-modal Database for Answering Complex Natural Language Queries}

\author{%
  Mahdi Erfanian\\
  Department of Computer Science\\
  University of Illinois Chicago\\
  Chicago, IL \\
  \texttt{merfan2@uic.edu} \\
  \And
  Mohsen Dehghankar\\
  Department of Computer Science\\
  University of Illinois Chicago\\
  Chicago, IL \\
  \texttt{mdehgh2@uic.edu} \\
  \And
  Abolfazl Asudeh \\
  Department of Computer Science \\
  University of Illinois Chicago \\
  Chicago, IL \\
  \texttt{asudeh@uic.edu} \\
}

\begin{document}

\maketitle

\begin{abstract}
Multi-modal datasets, like those involving images, often miss the detailed descriptions that properly capture the rich information encoded in each item. This makes answering complex natural language queries a major challenge in this domain. In particular, unlike the traditional nearest neighbor search, where the tuples and the query are represented as points in a single metric space, these settings involve queries and tuples embedded in fundamentally different spaces, making the traditional query answering methods inapplicable. Existing literature addresses this challenge for image datasets through vector representations jointly trained on natural language and images. This technique, however, underperforms for complex queries due to various reasons.

This paper takes a step towards addressing this challenge by introducing a Generative-based Monte Carlo method that utilizes foundation models to generate synthetic samples that capture the complexity of the natural language query and represent it in the same metric space as the multi-modal data.

Following this method, we propose \system, a database for image data retrieval. 
Instead of relying on contrastive learning or metadata-searching approaches, our system is based on synthetic data generation to capture the complexities of natural language queries. Our system is open-source and ready for deployment, designed to be easily adopted by researchers and developers.
The comprehensive experiments on various benchmark datasets verify that this system significantly outperforms state-of-the-art text-to-image retrieval methods in the literature.
Any foundation model and embedder can be easily integrated into \system to improve the performance, piggybacking on the advancements in these technologies.
\end{abstract}

\section{Introduction}\label{sec:intro}

Multi-modal datasets, like images, pose new challenges for data management systems.
Unlike traditional databases, where tuple attributes are explicitly specified, these datasets often miss proper descriptions that capture the rich information encoded in each item.
As a result, traditional query-answering approaches are not directly applicable in these settings.
Meanwhile, modern personal devices like smartphones have made it possible to collect large amounts of multi-modal data, even from individual users.
Due to the wealth of information ``hidden'' in each multi-modal tuple, the users may compose {\em complex natural language queries}, searching for the tuple(s) they are interested in.
To further motivate this, let us consider the following running example.

\begin{exbox}\label{ex:1}
{\small
    A photo enthusiast has collected a large pool of images over the past several years in her private data repository.
    She wants to retrieve a specific photo she has in mind. She describes the picture as \hl{\tt [A banana gazing at its reflection in a mirror]}.\footnotemark
}
\end{exbox}\footnotetext{See Figure~\ref{fig:specific_queries} for more examples.}

\begin{wrapfigure}{r}{0.5\textwidth}
  \begin{subfigure}[t]{\linewidth} 
    \vspace{-4mm}
    \includegraphics[width=\linewidth]{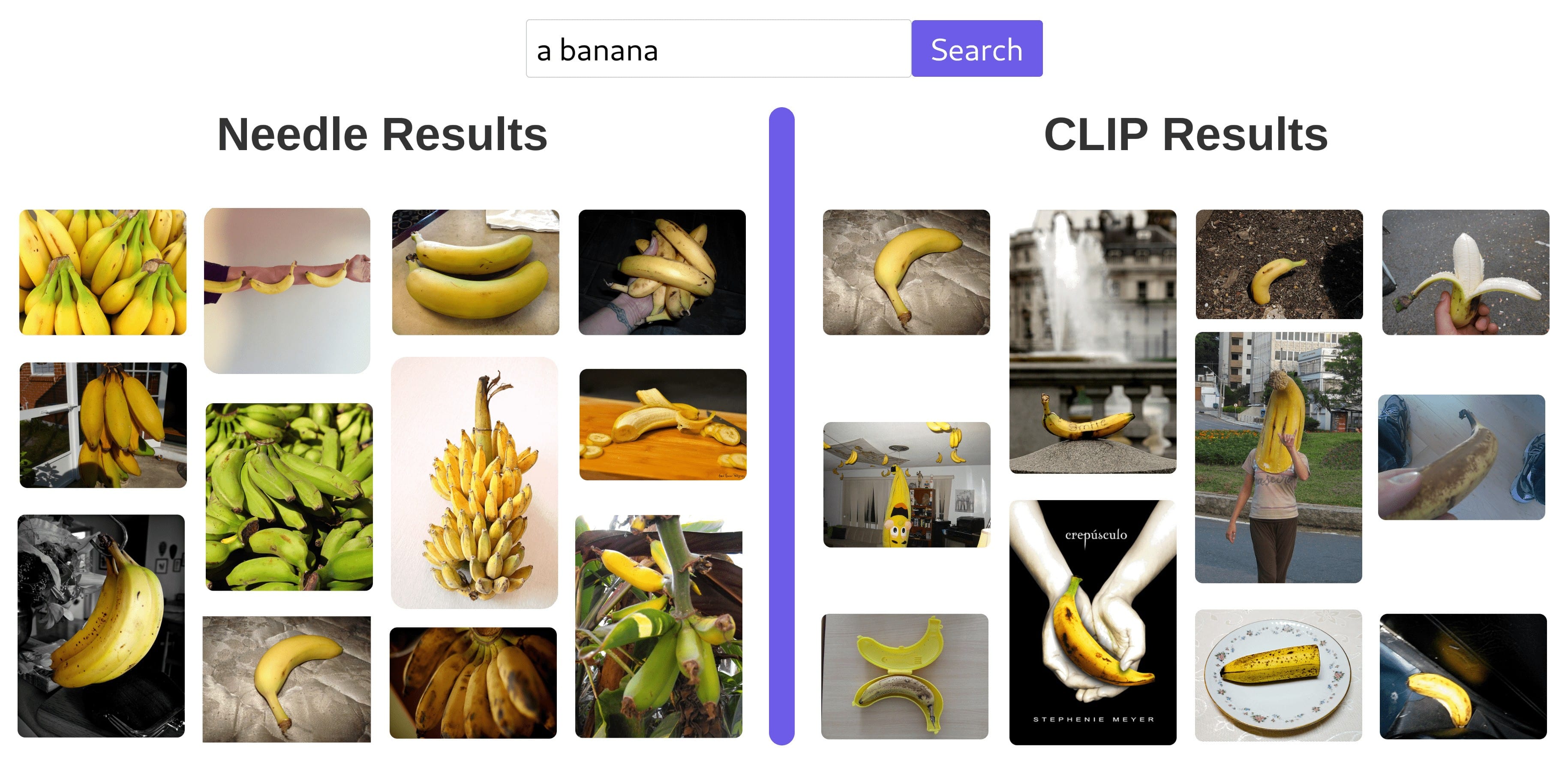}
    \tikz\draw[gray, dashed, thick](0,0)--(\linewidth,0);
  \end{subfigure} \\ 
  \begin{subfigure}[t]{\linewidth}
    \includegraphics[width=\linewidth]{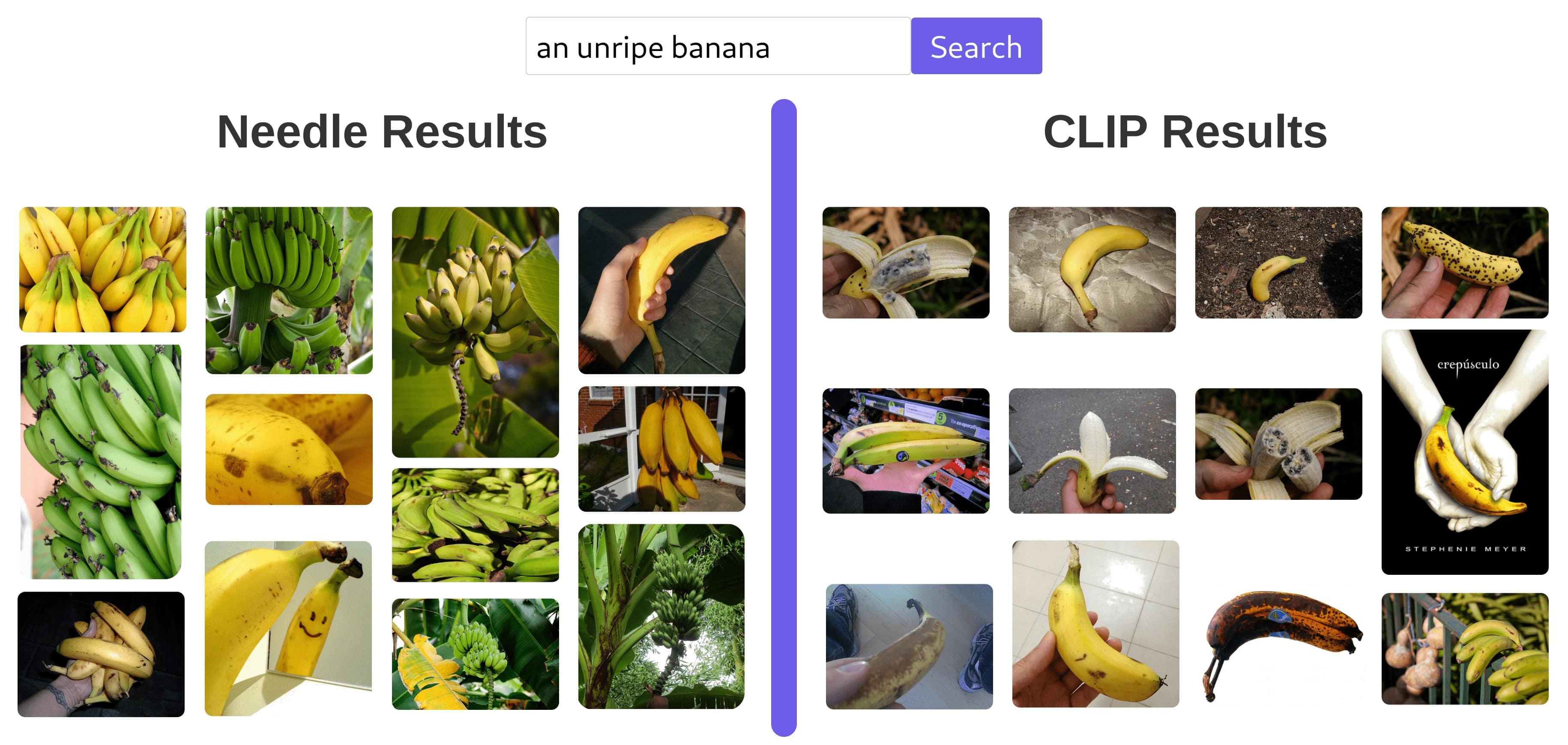}
    \tikz\draw[gray, dashed, thick](0,0)--(\linewidth,0);
  \end{subfigure} \\ 
  \begin{subfigure}[t]{\linewidth} 
    \includegraphics[width=\linewidth]{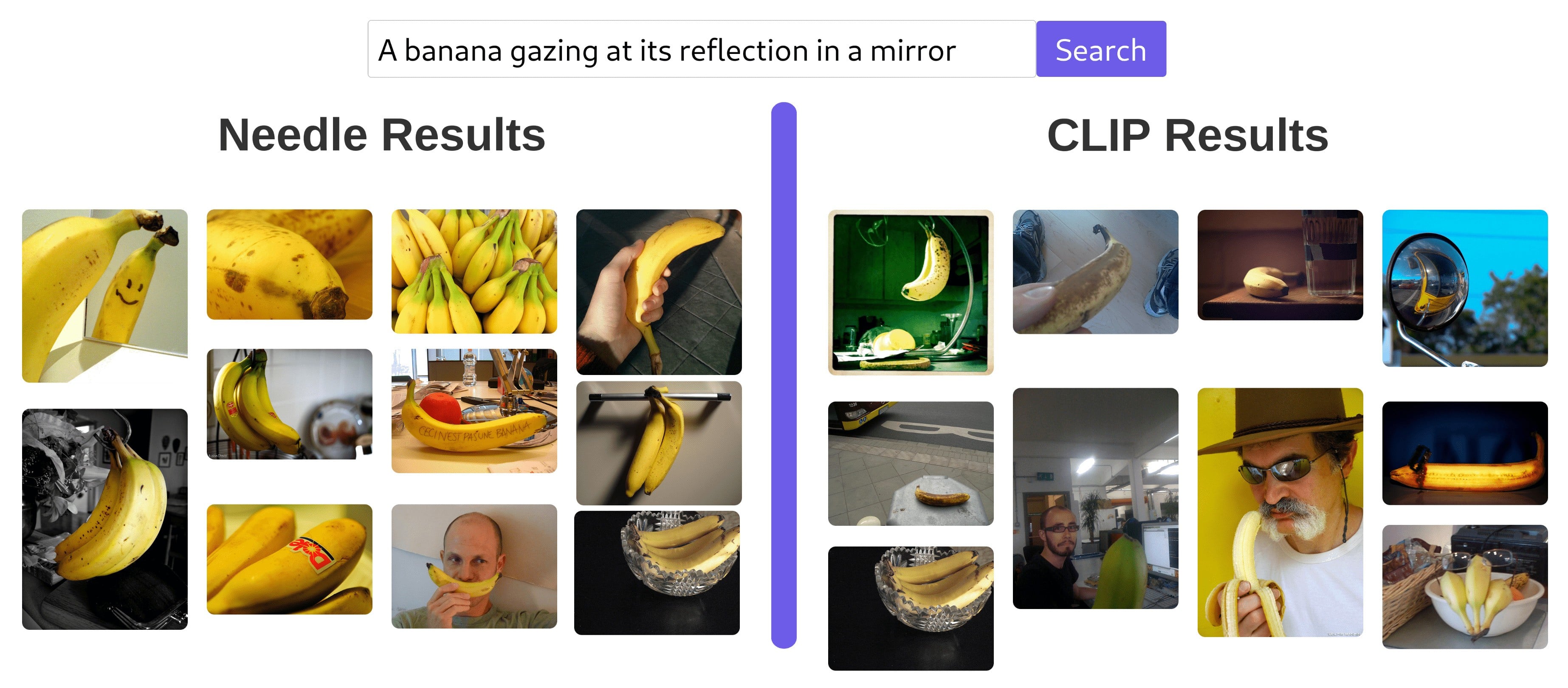}
  \end{subfigure}
  \vspace{-4mm}
  \caption{The query results for (a) [a banana] (row 1), (b) [an unripe banana] (row 2), (c) [a banana gazing at its reflection in a mirror] (row 3), using \system vs.\ \clip~\cite{radford2021learning}.}
  \label{fig:intro_example}
  \vspace{-7mm}
\end{wrapfigure}

Traditional Nearest Neighbor (NN) search algorithms are not directly applicable here since they consider the dataset tuples and the query as points in the same data space. In contrast, in Example~\ref{ex:1}, (i) each tuple is an image, and (ii) the query is in the form of an unstructured, complex natural language statement. 

The state-of-the-art approaches for answering natural language queries on image datasets train a vector representation (aka embedding) jointly on hundreds of millions of $\langle$image, text$\rangle$ pairs.
Using a jointly trained embedding, one can transform the query and the dataset images to vector representations in the same space and apply NN-search to answer the query.
This approach works satisfactorily for ``simple'' and ``common'' queries, such as simple object detection tasks. 

As a concrete example, continuing with our running example (Example~\ref{ex:1}), we use \clip~\cite{radford2021learning}, OpenAI’s well-known jointly-trained embedding model, to answer the natural language query \hl{\tt [A banana]} on the benchmark image dataset \lvis~\cite{gupta2019lvis}.
As reflected in the top-right (row 1, col. 2) cell of Figure~\ref{fig:intro_example}, the results are satisfactory, as most of the returned images include a banana. %
However, it fails for a moderately complex query, \hl{\tt [An unripe banana]}, as none of the returned images shows an unripe (green) banana -- the middle-right (row 2, col. 2) cell of Figure~\ref{fig:intro_example}.
It also fails on more complex queries, such as the one in Example~\ref{ex:1}, and does not retrieve the target image, the one shown in the bottom-right cell (row 3, col. 2) of Figure~\ref{fig:intro_example}.
As we further explain in \S~\ref{sec:exp}, we observe a similar behavior to this example across a wide range of queries.

This motivates us to take a step toward filling this gap by proposing a Monte Carlo randomized algorithm. In particular, we empower our algorithm with generative AI (GenAI) to generate a set of synthetic multi-modal tuples that represent the complex natural language query. 
Subsequently, leveraging a collection of embedders, we conduct a series of NN searches using the generated tuples and aggregate the results to generate the final query outputs.
We use our method to develop \system\footnote{The name \system is inspired by the common expression "finding \system in the haystack," reflecting the challenge of pinpointing relevant information in large multimodal spaces.}, our deployment-ready open-source system\footnote{
\system Github: \textcolor{blue}{\url{https://github.com/UIC-InDeXLab/Needle/tree/main/}}
} %
for complex natural query answering on image datasets. 

Unlike existing approaches, \system can handle natural language queries with different levels of complexity. 
To demonstrate this, let us consider our running example, Example~\ref{ex:1} once again. %
We issued the same set of simple, moderate, and complex queries to \system.
The results are provided in the first column of Figure~\ref{fig:intro_example}.
For the simple object-finding task, the top-left (row 1, col 1) cell in Figure~\ref{fig:intro_example}, we observed a similar (but slightly better) performance since all returned images satisfied the query, i.e., they included a banana.
It also performed reasonably well for the moderately complex query, as a large portion of the output images were green bananas -- the middle-left (row 2, col 1) cell in Figure~\ref{fig:intro_example}.
But perhaps, more interestingly, it could successfully find the image-of-interest described in the complex query of Example~\ref{ex:1} -- the bottom-left (row 3, col 1) cell in Figure~\ref{fig:intro_example}.
Interestingly, the target image was returned as {\em the top-1 search result} (the top-left image in the bottom-left cell), demonstrating the ability of our system to answer these types of queries.
We observed a consistent behavior for our system in experiments across a large number of different queries (\S~\ref{sec:exp}).

\stitle{Summary of contributions}
This paper introduces a new approach to using GenAI for complex natural-language query answering on multi-modal data. We propose \system, a system for image data retrieval.
 In summary, our contributions are the following:

\vspace{-2mm}
\begin{itemize}[leftmargin=*]
    \item We introduce a novel Monte Carlo method for answering complex natural-language queries on multi-modal data. We leverage Foundation models to generate synthetic tuples representing the queries. We then use a collection of vector representations (aka embeddings) that allows us to apply traditional k-NN techniques to retrieve related tuples. Our Monte Carlo algorithm aggregates the output generated for each (synthetic tuple, embedding) pair to obtain the final result. (\S~\ref{sec:method})
    
    \item We present \system, a concrete, interactive system built upon our Monte Carlo method. Developed with production readiness in mind, \system emphasizes ease of deployment, efficiency, robustness, and user-friendly interaction via a command-line interface (\S~\ref{sec:system}, Appendix~\ref{sec:system:dev}).
    \item We propose and integrate several practical optimizations designed to significantly improve \system's performance and efficiency, including a dynamic embedder trust mechanism, outlier detection for generated data, caching, and a query complexity classifier (\S~\ref{sec:prac}).
    \item We conduct a comprehensive experimental evaluation demonstrating \system's effectiveness across diverse benchmarks (including object detection and complex natural language queries) compared to state-of-the-art baselines, supported by hyperparameter analysis and a human study indicating strong user preference for \system (\S~\ref{sec:exp:poc:od}, \S~\ref{sec:exp:poc:nlq}, Appendix~\ref{sec:exp:ablation}, Appendix~\ref{sec:exp:study}).
\end{itemize}

\vspace{-4mm}
\section{Related Work}\label{sec:relatedwork}
\vspace{-1mm}

Our paper relates to the following literature:

\stitle{Multi-modal Data Retrieval}
Multi-modal data retrieval (aka cross-modal retrieval), integrating various data types, has recently attracted significant attention. Early methods focused on shared representations by mapping modalities into a common latent space using canonical correlation analysis (CCA)~\cite{rasiwasia2010new, wang2015cluster} and kernel-based techniques~\cite{jia2019semantically}. Recent approaches leverage deep learning models, including cross-modal encoders~\cite{wang2015image, liu2017learning} and joint embedding networks~\cite{balaneshin2018deep,rubio2017multi}. 
Advanced techniques, such as attention mechanisms~\cite{yang2021rethinking, tang2023interacting} and contrastive learning~\cite{pham2024composing}, further enhance retrieval by capturing fine-grained relationships. CLIP~\cite{radford2021learning} and ALIGN~\cite{jia2021scaling} are widely adopted methods for cross-modal retrieval, both of which are based on contrastive learning approaches.
Feedback mechanisms~\cite{moll2023seesaw, zhou2003relevance} and user preferences~\cite{zhang2010interactive, chen2020image} improve multi-modal search results. In addition, there exists practical approaches to enrich the semantic metadata of images using image tagging~\cite{fu2015tagging, zhang2024recognize, li2022blip} and object detection techniques~\cite{zou2023object}. Research also extends to other modalities, such as text-video retrieval~\cite{li2019w2vv++, markatopoulou2017query, torabi2016learning, mithun2018learning, miech2019howto100m, luo2021clip4clip} and text-audio retrieval~\cite{xin2023improving, chechik2008large, koepke2022audio}. For a detailed survey of cross-modal retrieval methods and future directions, see~\cite{li2023cross,wang2025cross}.

To the best of our knowledge, this paper is {\em the first} to introduce the idea of utilizing GenAI to synthetically generate samples that capture natural language complexities and proposes a GenAI-powered Monte Carlo method for multi-modal data retrieval.

\stitle{GenAI in Data Management}
The intersection of Generative AI (GenAI) and data management has emerged as a promising area, driving innovations in how data is generated, managed, and utilized. GenAI models, such as GPT~\cite{achiam2023gpt} and Stable Diffusion~\cite{rombach2022high}, have been employed for synthetic data generation to augment datasets~\cite{tanaka2019data,shahbazi2024coverage}, improve model training~\cite{le2017using, tremblay2018training}, and address data scarcity or imbalance issues~\cite{erfanian2024chameleon}. In data cleaning and integration, these models assist with tasks like imputation~\cite{villaizan2024diffusion}, and anomaly detection by generating plausible data patterns~\cite{ye2024saliencycut}. Recent works also explore the role of GenAI in automating query generation~\cite{trummer2022codexdb, trummer2023demonstrating, li2024using} and data summarization~\cite{dar2024advanced}, enhancing the efficiency of database systems and analytics platforms.

\stitle{Natural Language Query Processing}
The evolution of natural language to structured query translation has advanced from rule-based systems to sophisticated large language models (LLMs). Early methods used template matching and predefined rules~\cite{androutsopoulos1995natural, li2014nalir}, which were inflexible and required frequent manual updates. Machine learning approaches introduced more adaptability by learning query patterns from annotated datasets~\cite{ma2020mention, zhang2019editing}, though they struggled with complex language constructs.
Deep learning models, especially transformers like BERT~\cite{devlin2018bert} and GPT~\cite{brown2020language}, enabled semantic understanding and laid the foundation for text-to-SQL systems~\cite{li2024can}. Modern LLMs, such as PaLM~\cite{chowdhery2023palm} and GPT-4~\cite{achiam2023gpt}, generate context-aware queries across multiple domains with minimal supervision. Retrieval-augmented generation (RAG) further enhances query precision by leveraging external knowledge~\cite{lewis2020retrieval}. Ongoing efforts focus on fine-tuning LLMs to improve accuracy and efficiency in converting natural language to structured queries~\cite{trummer2022codexdb, trummer2023demonstrating, li2024using}.

\vspace{-3mm}
\section{The GenAI-powered Monte-Carlo Method}\label{sec:method}

We consider the dataset $\dee = \{t_1, t_2, \ldots, t_n\}$ as a set of $n$ multi-modal tuples with no explicit values on specific attributes or meta-data description about them.
This model complies with real-world needs, where personal devices such as cell phones have advanced cameras, enabling a vast volume of multi-modal data collection and sharing by any user of such devices.

The input query $\nlq$ is defined as a natural language description that corresponds to a subset of tuples $\dee_\nlq \subseteq \dee$, where $\dee_\nlq = \{t'_1, \cdots, t'_k \}$ represents the target subset for retrieval.

\stitle{Nearest-Neighbor Search}
Nearest-neighbor search in traditional settings is a well-studied line of research where many algorithms and indices such as Voronoi diagrams~\cite{kolahdouzan2004voronoi}, tree data structures~\cite{yianilos1993data}, and local sensitive hashing~\cite{jafari2021survey, huang2015query} have been proposed for efficient exact and approximate query answering.
In such settings, the data points are specified as points in a $d$ dimensional space, i.e., $t_i\in\Re^d$, where the goal is to find the $k$ nearest neighbors to a query point $q \in \Re^d$ based on a distance function $dist: \Re^d\times \Re^d\rightarrow \Re$.
That is, $k\mbox{-min}_{t_i\in\dee} dist(q,t_i)$. 
Such approaches, however, are not directly applicable in our setting because of two main challenges. (a) Multi-modal tuples such as images are not represented as points in $\Re^d$. and (b) The query is in the form of a (possibly complex) natural language (NL) phrase.
Fortunately, vector representations (aka embeddings)~\cite{oquab2014learning, kiela2014learning} have been proposed to address the first challenge. The state-of-the-art resolution to the second challenge is by utilizing the embeddings jointly trained on $\langle$tuple, text$\rangle$ pairs -- which enables transforming the query and the tuples into the same embedding space and applying NN-search to answer the query. However, as we observe in \S~\ref{sec:exp}, while this idea works well for simple queries such as object detection, it fails for more complex queries similar to the one in Example~\ref{ex:1}.
Therefore, we instead propose a new idea by {\em leveraging recent advancements in Generative AI and developing a randomized Monte Carlo algorithm} for (approximate) nearest-neighbor search in our setting.

\stitle{Vector Representations} %
A vector representation $\embedding: dom(t)\rightarrow \Re^{d_\mathcal{E}}$ is a transformation of the tuples to a high-dimensional vector of numeric values where the semantic similarity of tuples is proportional to the cosine similarity of their vector representations.
Specifically, given a vector representation (aka {\em embedder}) $\embedding^\ell$, we present $\embedding^\ell(t_i)$ as $\vec{v}^\ell_i =  \langle v_1, v_2, \cdots, v_{d_{\embedding^\ell}} \rangle$.
The distance between two vector embeddings $\vec{v}^\ell_i$ and $\vec{v}^\ell_j$ is computed based on their cosine similarity as

\vspace{-4mm}
\begin{align}
    \delta^\ell_{i,j} = \delta(\vec{v}^\ell_i,\vec{v}^\ell_j) = 1 - \cos\left(\angle(\vec{v}^\ell_i, \vec{v}^\ell_j)\right)
\end{align}

\vspace{-2mm}
Let $\Delta(t_i,t_j)$ be the (unknown) semantic distance of the tuples $t_i$ and $t_j$.
A vector representation $\embedding^o$ is optimal if it captures the semantic similarity of tuples, i.e., \(\delta^o_{i,j} \sim \Delta(t_i,t_j),~\forall t_i,t_j\).
Formally, 

\vspace{-4mm}
\begin{align}\label{eq:optimalEmbedding}
 \Delta(t_i,t_j)\geq \Delta(t_i,t_k)
    ~\Leftrightarrow~ \delta^o_{i,j}\geq \delta^o_{i,k}
\end{align}

\vspace{-2mm}
In practice, however, vector representations are learned using a universe of available multi-modal data.
As a result, instead of guaranteeing Equation~\ref{eq:optimalEmbedding}, the learned embeddings satisfy a (weaker) guarantee in expectation. 
That is, the {\em expected} distance between the embeddings of two tuples is proportional to their semantic distance, i.e., \(\ee\left[\delta^\ell_{i,j}\right] \sim \Delta(t_i,t_j),~\forall t_i,t_j\).

\stitle{A Baseline Approach} %
A vector representation that is jointly trained on the multi-modal data and textual data transform text and image to the vector-representation space. Therefore, given a natural language query, one can compute its embedding and find its nearest neighbor images in the embedding space.
This approach, however, suffers from two major issues:

First, the paired texts with the images are usually simple sentences (e.g., \at{[A photo of a dog]}), unable to fully describe the rich information encoded in the image. Conversely, our goal is to answer complex natural language queries, describing a scene with multiple objects and their relationships (e.g., \at{[A dog sitting by the fireplace, staring at the cat on the couch]}).
As a result, as we shall verify in our experiments (\S~\ref{sec:exp}), 
while such embeddings may perform reasonably on simple natural language queries, their performance drops significantly on complex queries.

Second, as explained earlier, learned embeddings only preserve semantic similarities in expectation.
As a result, relying solely on one vector representation for query answering may be misleading and inaccurate.
Hence, we instead utilize an ensemble (a set) of vector representation transformations, possibly trained on different training datasets, to develop our Monte Carlo algorithm for more accurate query answering.

\vspace{-3mm}
\subsection{Query Transformation using GenAI}\label{sec:method:transformation}
\vspace{-1mm}

In traditional nearest-neighbor problems, both the dataset tuples and the query are represented as points in the same space.
However,  in our problem, the query is in the form of a natural language phrase, while the dataset contains multi-modal tuples.

Let $\nlq$ be the given natural language query, and let $\{\vec{v}^o_1,\cdots, \vec{v}^o_n\}$ be the vector representations for the tuples in $\dee$.
If we could specify the corresponding vector representation $\vec{q}^o(\nlq)$  for $\nlq$ in the $\embedding^o$ space, we could find the $k$-nearest neighbors of $\nlq$ as 

\vspace{-10mm}
\begin{align}\label{eq:NN}
   \hspace{60mm} k\mbox{-NN}(\nlq,\dee) = \underset{t_i\in\dee}{k\mbox{-argmin}}~\delta\left(\vec{q}^o(\nlq),\vec{v}^o_i\right)
\end{align}

\vspace{-3mm}
As a result, if we could generate the synthetic multi-modal tuple $\gee_\nlq$ for which $\embedding^o(\gee_\nlq) = \vec{q}^o(\nlq)$, we could use Equation~\ref{eq:NN} for query answering.

Fortunately, recent advancements in generative AI, particularly the emergence of the foundation models (e.g., \dalle) based on guided-diffusion techniques~\cite{zhang2023text, rombach2022high, ramesh2022hierarchical}, have enabled the generation of synthetic multi-modal tuples for complex prompts $p$ (natural language descriptions of the tuple to be generated).
However, due to many reasons such as the randomized nature of the foundation models, the generated tuples\footnote{Throughout this paper, we refer to these synthetic tuples as ``{\bf guide tuples}'' that represent the natural language query in search space. We use these tuples to \textit{``guide''} our system toward retrieving the most relevant results.} may not be the optimal synthetic tuple $\gee_\nlq$, but we can assume that the generated tuple is (semantically) in the vicinity of $\gee_\nlq$.

Formally, we assume every tuple generated by the foundation model for the prompt $p=toPrompt(\nlq)$ lies randomly within a ($d_{\dist^o}$-dimensional) ball centered at $\gee_{\nlq}$ with the radius $\eps$. Therefore, for a sample tuple $\bar{\gee}_{\nlq}$ we have,

\vspace{-4mm}
\[
    \mathbb{E}\left[\embedding^o(\bar{\gee}_{\nlq})\right] = \embedding^o(\gee_{\nlq})
\]
\vspace{-2mm}

The final result follows from the principle that if the expected value of the cosine similarity between two vectors is 1, then the vectors are aligned in expectation.

\subsection{The Randomized Algorithm}

We are now ready to develop our Monte Carlo randomized algorithm based on repeated sampling.

Given a natural language query $\nlq$, and at least one foundation model $\mathcal{F}$, at the first step, we generate $m$ synthetic guide tuples $ \{\bar{\gee}_1,\cdots, \bar{\gee}_m\}$. As mentioned before, we consider each guide tuple $\bar{\gee}_i$ as an iid (independent and identically distributed) sample from a distribution with mean $\gee_\nlq$.
Furthermore, as mentioned earlier, instead of relying on the distance estimations based on one vector representation, we view each embedder $\embedding^{\ell}$ as a sample from the space possible embedders. Hence, we use a set $\{\embedding^1, \cdots, \embedding^l\}$ of embedders to minimize the semantic distance estimation error.
For each embedder $\embedding^\ell$, we assume
\(\ee\left[\delta^\ell_{i,j}\right] = \delta^o_{i,j},~\forall t_i,t_j\).

Using the generated tuples $\{\bar{\gee}_1,\cdots, \bar{\gee}_m\}$ and the embedders $\{\embedding^1, \cdots, \embedding^l\}$, we estimate the distance of each tuple $t_i$ to the query $\nlq$ as

\vspace{-10mm}
\begin{align}\label{eq:distanceEstimation}
    \hspace{20mm} \bar{\delta}_{\nlq,i} = \frac{1}{m\, l}\sum_{j=1}^m\sum_{\ell=1}^l ~\delta\left( \embedding^\ell(\bar{\gee}_j), \vec{v}^\ell_i\right)
\end{align}

We propose following theorem to show that with a high probability the distance estimations by Equation~\ref{eq:distanceEstimation} are near-optimal.

\begin{theorem}\label{th:errorbound}
    \footnote{The proof is provided in Appendix~\ref{proof:th1}.}
    Let \(\delta_{\nlq,i} = \delta\left(\vec{q}(\nlq),\vec{v}^o_i\right)\), $\forall t_i\in \dee$.
    Given a positive value $\gamma$,
    \[
    \Pr\left(
    \left(\frac{\bar{\delta}_{\nlq,i}}{\delta_{\nlq,i}}\geq (1+\gamma)\right) \vee 
    \left(\frac{\bar{\delta}_{\nlq,i}}{\delta_{\nlq,i}}\leq (1-\gamma)\right)
    \right) \leq \mathbf{e}^{\frac{-m\,l\,\gamma^2\delta_{\nlq,i}}{3}} + \mathbf{e}^{\frac{-m\,l\,\gamma^2\delta_{\nlq,i}}{2}}.
    \]
\end{theorem}

Following Theorem~\ref{th:errorbound}, we can use the estimated distances to identify the $k$-nearest neighbors of $\nlq$ as
\[
\overline{k\mbox{-NN}}(\nlq,\dee) = \underset{t_i\in\dee}{k\mbox{-argmin}}~\bar{\delta}_{\nlq,i}
\]

\vspace{-9mm}
\subsection{Practical Optimizations}\label{sec:prac}
In addition to the theoretical results, we implement several practical optimizations to enhance our system's performance and efficiency. Due to space constraints, we provide a concise summary here, with full details elaborated in Appendix~\ref{sec:app:prac}. 
First, we consider aggregating the rankings based on embedders' trust scores, which are learned on the fly. Since embedders are trained on different datasets and employ different architectures, they may perform differently in practice.
Therefore, we assign a performance (trust) weight to each embedder conditioned on each topic. We propose an algorithm that dynamically adjusts the embedders' weights based on their performance across various topics using users' feedback, thereby optimizing their contribution to the final aggregated ranking (Appendix~\ref{sec:prac:agg}). To improve the robustness, we incorporate an outlier detection mechanism to discard potentially misaligned or corrupted generated images from the guide set (Appendix~\ref{sec:prac:anomaly}). Furthermore, we enhance inference speed through two parallel mechanisms: a caching system that leverages the previously issued queries and acquired metadata to expedite similar query responses (Appendix~\ref{sec:prac:metadata}), and a query classifier that distinguishes simple from hard queries, allowing direct retrieval for simple cases without requiring guide image generation (Appendix~\ref{sec:prac:complexity}).

\vspace{-4mm}
\section{System Overview}\label{sec:system}
\vspace{-1mm}

\begin{figure*}
    \centering
    \includegraphics[width=\textwidth]{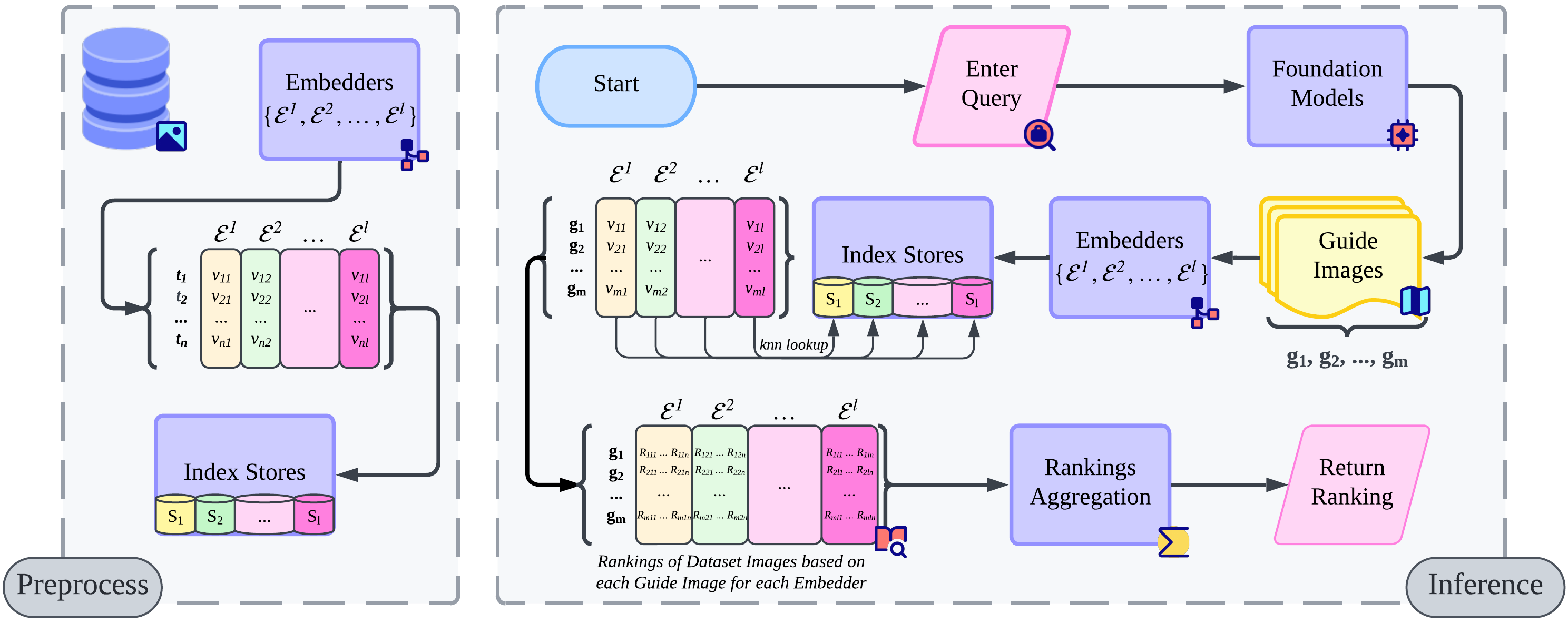}
    \vspace{-6mm}
    \caption{\system Architecture}
    \label{fig:needle_arch}
    \vspace{-6mm}
\end{figure*}

In the preceding section, we presented the theoretical foundations of our approach. In this section, we detail how these principles form the design of our image content retrieval system. Although our GenAI-powered Monte Carlo method is modality-agnostic, our focus here is on its application to image retrieval. Figure~\ref{fig:needle_arch} illustrates the overall architecture of \system, which is organized into two primary components: {\em Preprocessing} and {\em Inference}.\footnote{System development details is provided in Appendix~\ref{sec:system:dev}}

\stitle{Preprocessing}\label{sec:system:preprocessing}
The preprocessing phase involves a one-time operation to create vector representations of the input dataset using a set of predefined embedders $\{\embedding^1, \cdots, \embedding^l\}$. Once the raw image data is processed into vectors, it is indexed within the vector store. The vector store provides a maximum inner product search interface with low latency, which is crucial since users expect prompt results from the system.

\stitle{Inference}\label{sec:system:inference}
At the query time, we need to generate guide tuples representing the query using foundation models. \system can use multiple (black-box) foundation models to generate guide images, with the number of guide images fine-tuned as a hyperparameter ($m$). Different foundation models might produce images with varying levels of quality and realism. However, for the purposes of \system, high-quality, realistic images are unnecessary to achieve accurate search results. As a result, lower-quality images are sufficient, as long as they closely resemble the input query. In our extended experiments in Appendix~\ref{sec:exp:ablation}, we will explore the impact of different configurations in the image generation process, like the size of the generated images, the number of guide images, the count, and the overall quality of the foundation models in the final results.

After generating the guide images, \system uses predefined embedders to transform guide tuples into vector representations and conducts a k-NN lookup for each generated image across all embedders. The results are then aggregated into a final ranking, where each embedder's contribution is calculated according to its current weight. This weight reflects the system's trust in the performance of the corresponding embedder.
While dynamic adjustment of these weights based on user feedback is supported (Appendix~\ref{sec:prac:agg}), the default configuration assumes such feedback is unavailable and uses fixed predefined weights, which can be derived from external performance indicators like timm leaderboard scores\footnote{\href{https://huggingface.co/spaces/timm/leaderboard}{https://huggingface.co/spaces/timm/leaderboard}}. This weighting mechanism is general and fully configurable by the user.

\section{Experimental Evaluation}\label{sec:exp}

In this section, we evaluate several key aspects of \system to demonstrate its capabilities and to evaluate its performance. 
We use multiple benchmark datasets and several baselines for this purpose.
In the following, we first detail our experiments setup (\S~\ref{sec:exp:setup}), followed by a proof-of-concept analysis (\S~\ref{sec:exp:poc}) that demonstrates \system's efficacy in text-to-image retrieval across diverse object detection (\S~\ref{sec:exp:poc:od}) and complex natural language query benchmarks (\S~\ref{sec:exp:poc:nlq}). Subsequently, we conduct an ablation study in Appendix~\ref{sec:exp:ablation} to assess the impact of hyperparameter variations on \system's performance. Finally, in Appendix~\ref{sec:exp:study} we provide an end-to-end case study, involving human participants' qualitative feedback on \system's responsiveness to arbitrary user-generated queries.

\vspace{-4mm}
\subsection{Experiments Setup}\label{sec:exp:setup}
\vspace{-1mm}

\stitle{Baselines}\label{sec:exp:setup:baselines}
We evaluate our method using several prominent vision-language models as baselines, representing different architectural and training paradigms. These include well-established contrastive learning models such as CLIP~\cite{radford2021learning} and ALIGN~\cite{jia2021scaling}, the unified multimodal approach of FLAVA~\cite{singh2022flava}, and the effective decoupled pipeline combining BLIP for captioning with MiniLM for text embedding~\cite{li2022blip,wang2020minilm}. Each of these models offers a unique perspective on learning joint image-text representations, and together they provide a comprehensive set of comparisons for our work. Detailed descriptions of these models and their specific configurations used in our experiments are deferred to Appendix~\ref{sec:appendix_models}.

\stitle{Datasets}\label{sec:exp:setup:datasets}
We evaluate \system using a comprehensive set of datasets, categorized into those focusing on object detection and those involving complex natural language queries. The object detection benchmarks include Caltech256~\cite{griffin_holub_perona_2022}, MS COCO~\cite{lin2014microsoft}, LVIS~\cite{gupta2019lvis}, and BDD100k~\cite{yu2020bdd100k}. For complex natural language queries, we utilize COLA~\cite{ray2023cola}, Winoground~\cite{thrush2022winoground}, NoCaps~\cite{agrawal2019nocaps}, and SentiCap~\cite{mathews2016senticap}. These datasets collectively provide diverse challenges, from identifying specific objects to understanding nuanced language and handling novel concepts or sentiments, essential for benchmarking zero-shot retrieval performance. Detailed descriptions of the datasets are provided in Appendix~\ref{sec:appendix_datasets}.

\stitle{Evaluation Metrics}\label{sec:exp:setup:metrics} 
We employ distinct evaluation metrics tailored to the characteristics of our two benchmark types: object detection and complex natural language queries. For object detection benchmarks, we measure performance using Mean Recall at $k$ (R@k), Precision at $k$ (P@k), Mean Average Precision (MAP), and Mean Reciprocal Rank (MRR). For complex natural language query experiments, we report both MRR and Pairing Accuracy (PAcc). Detailed definitions and formulas for these evaluation metrics are provided in Appendix~\ref{sec:appendix_metrics}.

\stitle{Embedders}\label{sec:exp:setup:embedders}
We use all or a subset of embedders listed in Appendix (Table~\ref{tab:embedder_weights}) in our experiments.

\stitle{Hardware Configuration} \label{sec:exp:setup:configuration}
All experiments were conducted on two servers, each equipped with 32 GB of memory, a 12-core CPU, and two Tesla T4 GPUs, running Ubuntu 22.04 LTS.

\stitle{Foundation Models and the Monetary Cost}\label{sec:exp:setup:fm}
For image generation in our experiments, we employed several state-of-the-art foundation models, namely \texttt{DALL-E}\footnote{\url{https://openai.com/index/dall-e-3/}}, \texttt{ImagenV3-fast}\footnote{\url{https://replicate.com/google/imagen-3-fast}}, \texttt{Flux-Schnell}\footnote{\url{https://replicate.com/black-forest-labs/flux-schnell}}, and \texttt{RealvisXL-v3.0-turbo}\footnote{\url{https://replicate.com/adirik/realvisxl-v3.0-turbo}}. These models were chosen for their complementary strengths in generating high-quality images. The total monetary cost incurred for utilizing these models was \$257.53 (USD).

\vspace{-3mm}
\subsection{Proof of Concept}\label{sec:exp:poc}
\vspace{-2mm}

\begin{figure*}[t!]
    \centering
\includegraphics[width=\textwidth]{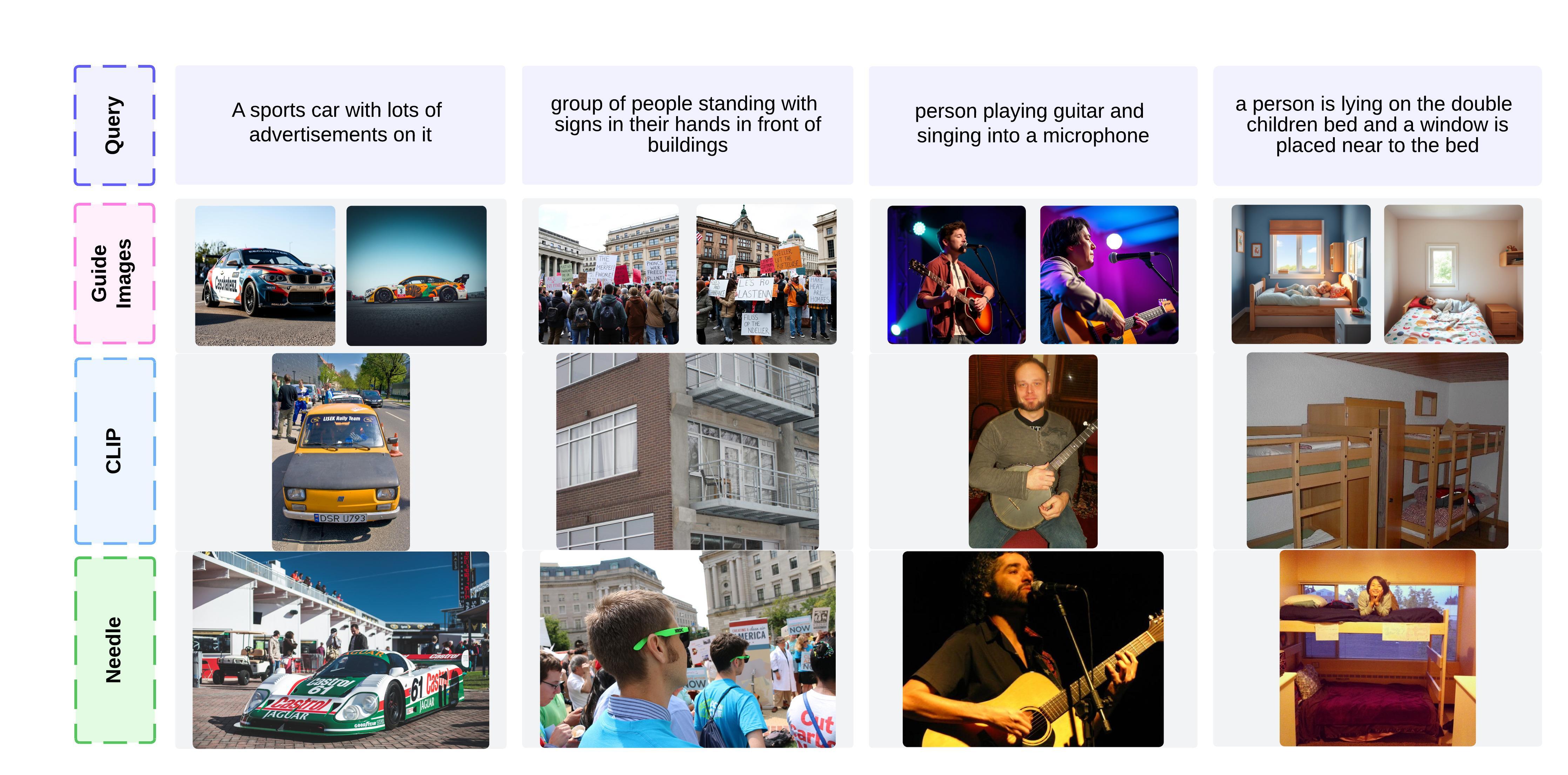}
\vspace{-6mm}
\caption{Illustration of Complex Natural Language Queries extracted from NoCaps~\cite{agrawal2019nocaps}, Guide images, and \clip vs. \system results.}
\label{fig:specific_queries}
\vspace{-5mm}
\end{figure*}

\subsubsection{{\sc Task 1:} Object Detection}\label{sec:exp:poc:od}

\begin{table}
    \centering
    \caption{Zero-shot retrieval performance on object detection datasets. For each dataset, we report R@10, P@10, MAP, and MRR. Each cell shows two numbers (All / \hard{Hard}).}
    \label{tab:od-results}
    \resizebox{\textwidth}{!}{%
    \begin{tabular}{lcccccccccccccccc}
        \toprule
         & \multicolumn{4}{c}{Caltech256} & \multicolumn{4}{c}{COCO} & \multicolumn{4}{c}{LVIS} & \multicolumn{4}{c}{BDD} \\
        \cmidrule(lr){2-5} \cmidrule(lr){6-9} \cmidrule(lr){10-13} \cmidrule(lr){14-17}
        Baseline & R@10 & P@10 & MAP & MRR & R@10 & P@10 & MAP & MRR & R@10 & P@10 & MAP & MRR & R@10 & P@10 & MAP & MRR \\
        \midrule
        CLIP            & \makecell{0.926 \\ \hard{0.150}}   & \makecell{0.926 \\ \hard{0.150}}   & \makecell{0.939 \\ \hard{0.181}}   & \makecell{0.952 \\ \hard{0.193}}   & \makecell{0.934 \\ \hard{0.400}}   & \makecell{0.934 \\ \hard{0.400}}   & \makecell{0.952 \\ \hard{0.477}}   & \makecell{\underline{0.988} \\ \hard{\textbf{1.000}}}   & \makecell{0.177 \\ \hard{0.093}} & \makecell{0.167 \\ \hard{0.083}} & \makecell{0.168 \\ \hard{0.078}} & \makecell{0.316 \\ \hard{0.224}} & \makecell{\underline{0.660} \\ \hard{\underline{0.033}}}   & \makecell{\underline{0.660} \\ \hard{\underline{0.033}}}   & \makecell{0.670 \\ \hard{0.005}}   & \makecell{0.714 \\ \hard{0.048}} \\
        ALIGN           & \makecell{\underline{0.941} \\ \hard{\underline{0.375}}}   & \makecell{\underline{0.941} \\ \hard{\underline{0.375}}}   & \makecell{\underline{0.947} \\ \hard{\underline{0.398}}}   & \makecell{\underline{0.961} \\ \hard{\underline{0.541}}}   & \makecell{\underline{0.944} \\ \hard{\underline{0.800}}}   & \makecell{\underline{0.944} \\ \hard{\underline{0.800}}}   & \makecell{\underline{0.960} \\ \hard{\underline{0.895}}}   & \makecell{0.981 \\ \hard{\textbf{1.000}}}   & \makecell{\underline{0.215} \\ \hard{\underline{0.145}}} & \makecell{\underline{0.201} \\ \hard{\underline{0.130}}} & \makecell{\underline{0.207} \\ \hard{\underline{0.129}}} & \makecell{\underline{0.379} \\ \hard{\underline{0.306}}} & \makecell{0.560 \\ \hard{0.000}}   & \makecell{0.560 \\ \hard{0.000}}   & \makecell{0.573 \\ \hard{0.003}}   & \makecell{0.704 \\ \hard{0.014}} \\
        FLAVA           & \makecell{0.882 \\ \hard{0.258}}   & \makecell{0.882 \\ \hard{0.258}}  & \makecell{0.903 \\ \hard{0.306}}   & \makecell{0.949 \\ \hard{0.491}}   & \makecell{0.924 \\ \hard{0.100}}   & \makecell{0.924 \\ \hard{0.100}}   & \makecell{0.941 \\ \hard{0.281}}   & \makecell{0.963 \\ \hard{\textbf{1.000}}}   & \makecell{0.185 \\ \hard{0.109}} & \makecell{0.172 \\ \hard{0.097}} & \makecell{0.180 \\ \hard{0.099}}  & \makecell{0.321 \\ \hard{0.241}} & \makecell{\underline{0.660} \\ \hard{\underline{0.033}}}   & \makecell{\underline{0.660} \\ \hard{\underline{0.033}}}   & \makecell{\underline{0.698} \\ \hard{0.036}}   & \makecell{\underline{0.725} \\ \hard{\underline{0.083}}} \\
        BLIP + MiniLM   & \makecell{0.826 \\ \hard{0.317}}   & \makecell{0.826 \\ \hard{0.317}}   & \makecell{0.838 \\ \hard{0.372}}   & \makecell{0.880 \\ \hard{0.408}}   & \makecell{0.941 \\ \hard{0.700}}   & \makecell{0.941 \\ \hard{0.700}}   & \makecell{0.951 \\ \hard{0.698}}   & \makecell{0.975 \\ \hard{\textbf{1.000}}}   & \makecell{0.180 \\ \hard{0.115}} & \makecell{0.177 \\ \hard{0.111}} & \makecell{0.179 \\ \hard{0.107}} & \makecell{0.332 \\ \hard{0.260}} & \makecell{0.600 \\ \hard{\textbf{0.167}}}   & \makecell{0.600 \\ \hard{\textbf{0.167}}}   & \makecell{0.610 \\ \hard{\underline{0.144}}}   & \makecell{\underline{0.725} \\ \hard{\textbf{0.333}}} \\
        \hdashline
        Needle          & \makecell{\textbf{0.962} \\ \hard{\textbf{0.667}}}   & \makecell{\textbf{0.962} \\ \hard{\textbf{0.667}}}   & \makecell{\textbf{0.966} \\ \hard{\textbf{0.687}}}  & \makecell{\textbf{0.979} \\ \hard{\textbf{0.776}}}   & \makecell{\textbf{0.966} \\ \hard{\textbf{0.900}}}   & \makecell{\textbf{0.966} \\ \hard{\textbf{0.900}}}   & \makecell{\textbf{0.977} \\ \hard{\textbf{0.981}}}   & \makecell{\textbf{1.000} \\ \hard{\textbf{1.000}}}   & \makecell{\textbf{0.330} \\ \hard{\textbf{0.263}}} & \makecell{\textbf{0.295} \\ \hard{\textbf{0.225}}} & \makecell{\textbf{0.323} \\ \hard{\textbf{0.249}}} & \makecell{\textbf{0.511} \\ \hard{\textbf{0.453}}} & \makecell{\textbf{0.720} \\ \hard{\textbf{0.167}}}   & \makecell{\textbf{0.720} \\ \hard{\textbf{0.167}}}   & \makecell{\textbf{0.711} \\ \hard{\textbf{0.158}}}   & \makecell{\textbf{0.750} \\ \hard{\textbf{0.333}}} \\
        \bottomrule
    \end{tabular}%
    }
    \vspace{-6mm}
\end{table}

We begin our experiments by evaluating the performance of \system against the baselines on object detection benchmarks, utilizing the Caltech256, COCO, LVIS, and BDD datasets. 
For each object in these benchmarks, we formulate queries to retrieve them. 
For \system, we employed three image generation engines (\texttt{Flux-Schnell}, \texttt{RealVisV3}, and \texttt{ImagenV3-fast}) to generate three images per engine, resulting in a total of nine guide images per query, all at MEDIUM quality. Additionally, we utilized the ensemble of embedders detailed in Table~\ref{tab:embedder_weights}. Following the methodology outlined in Appendix~\ref{sec:appendix_models}, we considered queries with a CLIP Average Precision below 0.5 as the ``hard set''. 

Table~\ref{tab:od-results} presents the performance of \system against the baselines across R@10, P@10, MAP, and MRR metrics. Notably, \system demonstrates superior performance compared to all baselines on both the easy and hard sets. Specifically, for the hard set, \system achieves MAP improvements of 73\%, 10\%, 93\%, and 10\% over the second-best baseline on Caltech256, COCO, LVIS, and BDD datasets, respectively. These results underscore the substantial advantage of using synthetic data for image retrieval over traditional contrastive learning and image-to-text approaches. 
Furthermore, Figure~\ref{fig:specific_queries} provides a visual illustration of the performance of \system (v.s. CLIP) and the synthetic guide images generated for each query.

\vspace{-2mm}
\subsubsection{{\sc Task 2:} Complex Natural Language Queries}\label{sec:exp:poc:nlq}

\begin{table}
    \centering
    \caption{Zero-shot retrieval performance on complex natural language queries benchmarks.}
    \label{tab:complex-results}
    \begin{tabular}{lcccccc}
        \toprule
         & \multicolumn{2}{c}{Cola} & \multicolumn{2}{c}{Winoground} & SentiCap & NoCaps \\
        \cmidrule(lr){2-3} \cmidrule(lr){4-5} \cmidrule(lr){6-6} \cmidrule(lr){7-7}
        Baseline & PAcc & MRR & PAcc & MRR & MRR & MRR\\
        \midrule
        CLIP          & 0.578 & 0.246 & 0.519 & 0.426 & 0.464 & 0.573 \\
        ALIGN         & 0.591 & 0.301 & 0.554 & \textbf{0.501} & \underline{0.555} & \underline{0.704}\\
        FLAVA         & \underline{0.615} & \underline{0.336} & \underline{0.574} & 0.482 & 0.546  & 0.658 \\
        BLIP + MiniLM & 0.449 & 0.195 & 0.485 & 0.330 & 0.331 & 0.398\\
        \hdashline
        Needle        & \textbf{0.631} & \textbf{0.352} & \textbf{0.593} & \underline{0.490} & \textbf{0.642} & \textbf{0.745} \\
        \bottomrule
    \end{tabular}
    \vspace{-6mm}
\end{table}

Experiments involving complex natural language queries evaluate the system's ability to handle challenging language and visual nuances across different retrieval tasks. We define two types of evaluations:

First, the {\it Pairing Task} assesses the system's capability to correctly match a caption to its corresponding image when presented with two highly similar images and two corresponding captions that differ only in subtle details. Given four possible image-caption pairings, random chance achieves 0.25 Pairing Accuracy (PAcc). This task is evaluated using the Winoground~\cite{thrush2022winoground} and COLA~\cite{ray2023cola} datasets (see Figure~\ref{fig:cola-example} for examples from COLA illustrating the challenges of this setup).

Second, the {\it Full-set Retrieval Task} measures the system's performance when retrieving a specific image from the entire dataset using a single, potentially complex, query. For this task, we report the Mean Reciprocal Rank (MRR). This task is evaluated on the Winoground~\cite{thrush2022winoground}, COLA~\cite{ray2023cola}, SentiCap~\cite{mathews2016senticap}, and NoCaps~\cite{agrawal2019nocaps} benchmarks, as they feature challenging captions that require nuanced understanding for successful retrieval from a large pool.

Table~\ref{tab:complex-results} presents the Pairing Accuracy and Mean Reciprocal Rank (MRR) metrics for different aforementioned baselines. As illustrated in the table, \system outperforms the baselines in both PAcc and MRR. It is important to note that current foundation models still exhibit limitations in generating images that are fully aligned with an input query. In particular, our investigation reveals that these models frequently fail to produce images with the correct compositional ordering.
We anticipate that further advancements in foundation models will further enlarge the performance gap between \system and the baselines.
It is also noteworthy that the BLIP+MiniLM baseline performed significantly worse than the other baselines. Initially, we anticipated that this baseline would serve as a strong competitor in handling complex natural language queries, given its design to capture compositional and relational attributes between objects in images; however, the empirical results did not corroborate these expectations. In contrast, FLAVA demonstrated exceptionally strong performance in these tests.

\vspace{-3mm}
\section{Conclusion}\label{sec:conclusion}
\vspace{-1mm}
In this paper, we took a step towards addressing the challenge of answering complex natural language queries in multi-modal data settings by proposing a Generative-AI Powered Monte Carlo method. Our proposed method utilizes foundation models to generate synthetic guide tuples that capture the complexities of the input query; then, we leverage these guide tuples to represent the natural language query in multi-modal data space and apply traditional nearest neighbor methods. 

Following this method, We developed \system, an open-source database dedicated to delivering robust, efficient, and high-performance image retrieval\footnote{Due to the space constraints, the limitations of our system are discussed in Appendix~\ref{sec:lim}.}.
Our comprehensive experiments demonstrated that our system significantly outperformed the state-of-the-art techniques in image retrieval.

While our proposed technique is not limited to image datasets, its development for other modalities, such as audio and video, requires foundation models that are currently not publicly available. Upon the availability of such models, we aim to extend the scope of \system to a wider range of modalities in our future work.
We appreciate and highly value future contributions from the open-source community—whether through feedback, bug reports, feature requests, or direct code contributions—to help us further enhance and refine the system. %

\clearpage
\bibliography{ref}

\begin{thebibliography}{10}

\bibitem{achiam2023gpt}
Josh Achiam, Steven Adler, Sandhini Agarwal, Lama Ahmad, Ilge Akkaya, Florencia~Leoni Aleman, Diogo Almeida, Janko Altenschmidt, Sam Altman, Shyamal Anadkat, et~al.
\newblock Gpt-4 technical report.
\newblock {\em arXiv preprint arXiv:2303.08774}, 2023.

\bibitem{agrawal2019nocaps}
Harsh Agrawal, Karan Desai, Yufei Wang, Xinlei Chen, Rishabh Jain, Mark Johnson, Dhruv Batra, Devi Parikh, Stefan Lee, and Peter Anderson.
\newblock nocaps: novel object captioning at scale.
\newblock In {\em Proceedings of the IEEE International Conference on Computer Vision}, pages 8948--8957, 2019.

\bibitem{androutsopoulos1995natural}
Ion Androutsopoulos, Graeme~D Ritchie, and Peter Thanisch.
\newblock Natural language interfaces to databases--an introduction.
\newblock {\em Natural language engineering}, 1(1):29--81, 1995.

\bibitem{arampatzis2011modeling}
Avi Arampatzis and Stephen Robertson.
\newblock Modeling score distributions in information retrieval.
\newblock {\em Information Retrieval}, 14:26--46, 2011.

\bibitem{arora2012multiplicative}
Sanjeev Arora, Elad Hazan, and Satyen Kale.
\newblock The multiplicative weights update method: a meta-algorithm and applications.
\newblock {\em Theory of computing}, 8(1):121--164, 2012.

\bibitem{balaneshin2018deep}
Saeid Balaneshin-kordan and Alexander Kotov.
\newblock Deep neural architecture for multi-modal retrieval based on joint embedding space for text and images.
\newblock In {\em Proceedings of the Eleventh ACM International Conference on Web Search and Data Mining}, pages 28--36, 2018.

\bibitem{bao2021beit}
Hangbo Bao, Li~Dong, Songhao Piao, and Furu Wei.
\newblock Beit: Bert pre-training of image transformers.
\newblock {\em arXiv preprint arXiv:2106.08254}, 2021.

\bibitem{breunig2000lof}
Markus~M Breunig, Hans-Peter Kriegel, Raymond~T Ng, and J{\"o}rg Sander.
\newblock Lof: identifying density-based local outliers.
\newblock In {\em Proceedings of the 2000 ACM SIGMOD international conference on Management of data}, pages 93--104, 2000.

\bibitem{brown2020language}
Tom~B Brown.
\newblock Language models are few-shot learners.
\newblock {\em arXiv preprint arXiv:2005.14165}, 2020.

\bibitem{chechik2008large}
Gal Chechik, Eugene Ie, Martin Rehn, Samy Bengio, and Dick Lyon.
\newblock Large-scale content-based audio retrieval from text queries.
\newblock In {\em Proceedings of the 1st ACM international conference on Multimedia information retrieval}, pages 105--112, 2008.

\bibitem{chen2020image}
Yanbei Chen, Shaogang Gong, and Loris Bazzani.
\newblock Image search with text feedback by visiolinguistic attention learning.
\newblock In {\em Proceedings of the IEEE/CVF Conference on Computer Vision and Pattern Recognition}, pages 3001--3011, 2020.

\bibitem{chowdhery2023palm}
Aakanksha Chowdhery, Sharan Narang, Jacob Devlin, Maarten Bosma, Gaurav Mishra, Adam Roberts, Paul Barham, Hyung~Won Chung, Charles Sutton, Sebastian Gehrmann, et~al.
\newblock Palm: Scaling language modeling with pathways.
\newblock {\em Journal of Machine Learning Research}, 24(240):1--113, 2023.

\bibitem{cummins2012inference}
Ronan Cummins.
\newblock On the inference of average precision from score distributions.
\newblock In {\em Proceedings of the 21st ACM international conference on Information and knowledge management}, pages 2435--2438, 2012.

\bibitem{dar2024advanced}
Zaema Dar, Muhammad Raheel, Usman Bokhari, Akhtar Jamil, Esraa~Mohammed Alazawi, and Alaa~Ali Hameed.
\newblock Advanced generative ai methods for academic text summarization.
\newblock In {\em 2024 IEEE 3rd International Conference on Computing and Machine Intelligence (ICMI)}, pages 1--7. IEEE, 2024.

\bibitem{devlin2018bert}
Jacob Devlin.
\newblock Bert: Pre-training of deep bidirectional transformers for language understanding.
\newblock {\em arXiv preprint arXiv:1810.04805}, 2018.

\bibitem{erfanian2024chameleon}
Mahdi Erfanian, HV~Jagadish, and Abolfazl Asudeh.
\newblock Chameleon: Foundation models for fairness-aware multi-modal data augmentation to enhance coverage of minorities.
\newblock {\em Proceedings of the VLDB Endowment}, 17(11):3470--3483, 2024.

\bibitem{fagin2001optimal}
Ronald Fagin, Amnon Lotem, and Moni Naor.
\newblock Optimal aggregation algorithms for middleware.
\newblock In {\em Proceedings of the twentieth ACM SIGMOD-SIGACT-SIGART symposium on Principles of database systems}, pages 102--113, 2001.

\bibitem{fang2023eva}
Yuxin Fang, Wen Wang, Binhui Xie, Quan Sun, Ledell Wu, Xinggang Wang, Tiejun Huang, Xinlong Wang, and Yue Cao.
\newblock Eva: Exploring the limits of masked visual representation learning at scale.
\newblock In {\em Proceedings of the IEEE/CVF conference on computer vision and pattern recognition}, pages 19358--19369, 2023.

\bibitem{faria2010learning}
Fabio~F Faria, Adriano Veloso, Humberto~M Almeida, Eduardo Valle, Ricardo da~S Torres, Marcos~A Gon{\c{c}}alves, and Wagner Meira~Jr.
\newblock Learning to rank for content-based image retrieval.
\newblock In {\em Proceedings of the international conference on Multimedia information retrieval}, pages 285--294, 2010.

\bibitem{fu2015tagging}
Jianlong Fu, Tao Mei, Kuiyuan Yang, Hanqing Lu, and Yong Rui.
\newblock Tagging personal photos with transfer deep learning.
\newblock In {\em Proceedings of the 24th International Conference on World Wide Web}, pages 344--354, 2015.

\bibitem{griffin_holub_perona_2022}
Gregory Griffin, Alex Holub, and Pietro Perona.
\newblock Caltech 256, Apr 2022.

\bibitem{gupta2019lvis}
Agrim Gupta, Piotr Dollar, and Ross Girshick.
\newblock Lvis: A dataset for large vocabulary instance segmentation.
\newblock In {\em Proceedings of the IEEE/CVF conference on computer vision and pattern recognition}, pages 5356--5364, 2019.

\bibitem{huang2015query}
Qiang Huang, Jianlin Feng, Yikai Zhang, Qiong Fang, and Wilfred Ng.
\newblock Query-aware locality-sensitive hashing for approximate nearest neighbor search.
\newblock {\em Proceedings of the VLDB Endowment}, 9(1):1--12, 2015.

\bibitem{jafari2021survey}
Omid Jafari, Preeti Maurya, Parth Nagarkar, Khandker~Mushfiqul Islam, and Chidambaram Crushev.
\newblock A survey on locality sensitive hashing algorithms and their applications.
\newblock {\em arXiv preprint arXiv:2102.08942}, 2021.

\bibitem{jia2021scaling}
Chao Jia, Yinfei Yang, Ye~Xia, Yi-Ting Chen, Zarana Parekh, Hieu Pham, Quoc Le, Yun-Hsuan Sung, Zhen Li, and Tom Duerig.
\newblock Scaling up visual and vision-language representation learning with noisy text supervision.
\newblock In {\em International conference on machine learning}, pages 4904--4916. PMLR, 2021.

\bibitem{jia2019semantically}
Yuhua Jia, Liang Bai, Shuang Liu, Peng Wang, Jinlin Guo, and Yuxiang Xie.
\newblock Semantically-enhanced kernel canonical correlation analysis: a multi-label cross-modal retrieval.
\newblock {\em Multimedia Tools and Applications}, 78:13169--13188, 2019.

\bibitem{kiela2014learning}
Douwe Kiela and L{\'e}on Bottou.
\newblock Learning image embeddings using convolutional neural networks for improved multi-modal semantics.
\newblock In {\em Proceedings of the 2014 Conference on empirical methods in natural language processing (EMNLP)}, pages 36--45, 2014.

\bibitem{koepke2022audio}
A~Sophia Koepke, Andreea-Maria Oncescu, Jo{\~a}o~F Henriques, Zeynep Akata, and Samuel Albanie.
\newblock Audio retrieval with natural language queries: A benchmark study.
\newblock {\em IEEE Transactions on Multimedia}, 25:2675--2685, 2022.

\bibitem{kolahdouzan2004voronoi}
Mohammad Kolahdouzan and Cyrus Shahabi.
\newblock Voronoi-based k nearest neighbor search for spatial network databases.
\newblock In {\em Proceedings of the Thirtieth international conference on Very large data bases-Volume 30}, pages 840--851, 2004.

\bibitem{le2017using}
Tuan~Anh Le, Atilim~Giine{\c{s}} Baydin, Robert Zinkov, and Frank Wood.
\newblock Using synthetic data to train neural networks is model-based reasoning.
\newblock In {\em 2017 international joint conference on neural networks (IJCNN)}, pages 3514--3521. IEEE, 2017.

\bibitem{lewis2020retrieval}
Patrick Lewis, Ethan Perez, Aleksandra Piktus, Fabio Petroni, Vladimir Karpukhin, Naman Goyal, Heinrich K{\"u}ttler, Mike Lewis, Wen-tau Yih, Tim Rockt{\"a}schel, et~al.
\newblock Retrieval-augmented generation for knowledge-intensive nlp tasks.
\newblock {\em Advances in Neural Information Processing Systems}, 33:9459--9474, 2020.

\bibitem{li2014nalir}
Fei Li and Hosagrahar~V Jagadish.
\newblock Nalir: an interactive natural language interface for querying relational databases.
\newblock In {\em Proceedings of the 2014 ACM SIGMOD international conference on Management of data}, pages 709--712, 2014.

\bibitem{li2023cross}
Fengling Li, Lei Zhu, Tianshi Wang, Jingjing Li, Zheng Zhang, and Heng~Tao Shen.
\newblock Cross-modal retrieval: a systematic review of methods and future directions.
\newblock {\em arXiv preprint arXiv:2308.14263}, 2023.

\bibitem{li2024can}
Jinyang Li, Binyuan Hui, Ge~Qu, Jiaxi Yang, Binhua Li, Bowen Li, Bailin Wang, Bowen Qin, Ruiying Geng, Nan Huo, et~al.
\newblock Can llm already serve as a database interface? a big bench for large-scale database grounded text-to-sqls.
\newblock {\em Advances in Neural Information Processing Systems}, 36, 2024.

\bibitem{li2022blip}
Junnan Li, Dongxu Li, Caiming Xiong, and Steven Hoi.
\newblock Blip: Bootstrapping language-image pre-training for unified vision-language understanding and generation.
\newblock In {\em International conference on machine learning}, pages 12888--12900. PMLR, 2022.

\bibitem{li2019w2vv++}
Xirong Li, Chaoxi Xu, Gang Yang, Zhineng Chen, and Jianfeng Dong.
\newblock W2vv++ fully deep learning for ad-hoc video search.
\newblock In {\em Proceedings of the 27th ACM international conference on multimedia}, pages 1786--1794, 2019.

\bibitem{li2024using}
Zhenwen Li and Tao Xie.
\newblock Using llm to select the right sql query from candidates.
\newblock {\em arXiv preprint arXiv:2401.02115}, 2024.

\bibitem{lin2014microsoft}
Tsung-Yi Lin, Michael Maire, Serge Belongie, James Hays, Pietro Perona, Deva Ramanan, Piotr Doll{\'a}r, and C~Lawrence Zitnick.
\newblock Microsoft coco: Common objects in context.
\newblock In {\em Computer vision--ECCV 2014: 13th European conference, zurich, Switzerland, September 6-12, 2014, proceedings, part v 13}, pages 740--755. Springer, 2014.

\bibitem{liu2017learning}
Yu~Liu, Yanming Guo, Erwin~M Bakker, and Michael~S Lew.
\newblock Learning a recurrent residual fusion network for multimodal matching.
\newblock In {\em Proceedings of the IEEE international conference on computer vision}, pages 4107--4116, 2017.

\bibitem{luo2021clip4clip}
Huaishao Luo, Lei Ji, Ming Zhong, Yang Chen, Wen Lei, Nan Duan, and Tianrui Li.
\newblock Clip4clip: An empirical study of clip for end to end video clip retrieval.
\newblock {\em arXiv preprint arXiv:2104.08860}, 2021.

\bibitem{ma2020mention}
Jianqiang Ma, Zeyu Yan, Shuai Pang, Yang Zhang, and Jianping Shen.
\newblock Mention extraction and linking for sql query generation.
\newblock {\em arXiv preprint arXiv:2012.10074}, 2020.

\bibitem{malkov2018efficient}
Yu~A Malkov and Dmitry~A Yashunin.
\newblock Efficient and robust approximate nearest neighbor search using hierarchical navigable small world graphs.
\newblock {\em IEEE transactions on pattern analysis and machine intelligence}, 42(4):824--836, 2018.

\bibitem{markatopoulou2017query}
Foteini Markatopoulou, Damianos Galanopoulos, Vasileios Mezaris, and Ioannis Patras.
\newblock Query and keyframe representations for ad-hoc video search.
\newblock In {\em Proceedings of the 2017 ACM on international conference on multimedia retrieval}, pages 407--411, 2017.

\bibitem{mathews2016senticap}
Alexander Mathews, Lexing Xie, and Xuming He.
\newblock Senticap: Generating image descriptions with sentiments.
\newblock In {\em Proceedings of the AAAI conference on artificial intelligence}, volume~30, 2016.

\bibitem{mcinnes2018umap}
Leland McInnes, John Healy, and James Melville.
\newblock Umap: Uniform manifold approximation and projection for dimension reduction.
\newblock {\em arXiv preprint arXiv:1802.03426}, 2018.

\bibitem{miech2019howto100m}
Antoine Miech, Dimitri Zhukov, Jean-Baptiste Alayrac, Makarand Tapaswi, Ivan Laptev, and Josef Sivic.
\newblock Howto100m: Learning a text-video embedding by watching hundred million narrated video clips.
\newblock In {\em Proceedings of the IEEE/CVF international conference on computer vision}, pages 2630--2640, 2019.

\bibitem{mithun2018learning}
Niluthpol~Chowdhury Mithun, Juncheng Li, Florian Metze, and Amit~K Roy-Chowdhury.
\newblock Learning joint embedding with multimodal cues for cross-modal video-text retrieval.
\newblock In {\em Proceedings of the 2018 ACM on international conference on multimedia retrieval}, pages 19--27, 2018.

\bibitem{moll2023seesaw}
Oscar Moll, Manuel Favela, Samuel Madden, Vijay Gadepally, and Michael Cafarella.
\newblock Seesaw: interactive ad-hoc search over image databases.
\newblock {\em Proceedings of the ACM on Management of Data}, 1(4):1--26, 2023.

\bibitem{motwani1996randomized}
Rajeev Motwani and Prabhakar Raghavan.
\newblock Randomized algorithms.
\newblock {\em ACM Computing Surveys (CSUR)}, 28(1):33--37, 1996.

\bibitem{oquab2014learning}
Maxime Oquab, Leon Bottou, Ivan Laptev, and Josef Sivic.
\newblock Learning and transferring mid-level image representations using convolutional neural networks.
\newblock In {\em Proceedings of the IEEE conference on computer vision and pattern recognition}, pages 1717--1724, 2014.

\bibitem{oquab2023dinov2}
Maxime Oquab, Timoth{\'e}e Darcet, Th{\'e}o Moutakanni, Huy Vo, Marc Szafraniec, Vasil Khalidov, Pierre Fernandez, Daniel Haziza, Francisco Massa, Alaaeldin El-Nouby, et~al.
\newblock Dinov2: Learning robust visual features without supervision.
\newblock {\em arXiv preprint arXiv:2304.07193}, 2023.

\bibitem{pham2024composing}
Khoi Pham, Chuong Huynh, Ser-Nam Lim, and Abhinav Shrivastava.
\newblock Composing object relations and attributes for image-text matching.
\newblock In {\em Proceedings of the IEEE/CVF Conference on Computer Vision and Pattern Recognition}, pages 14354--14363, 2024.

\bibitem{radford2021learning}
Alec Radford, Jong~Wook Kim, Chris Hallacy, Aditya Ramesh, Gabriel Goh, Sandhini Agarwal, Girish Sastry, Amanda Askell, Pamela Mishkin, Jack Clark, et~al.
\newblock Learning transferable visual models from natural language supervision.
\newblock In {\em International conference on machine learning}, pages 8748--8763. PMLR, 2021.

\bibitem{ramesh2022hierarchical}
Aditya Ramesh, Prafulla Dhariwal, Alex Nichol, Casey Chu, and Mark Chen.
\newblock Hierarchical text-conditional image generation with clip latents.
\newblock {\em arXiv preprint arXiv:2204.06125}, 1(2):3, 2022.

\bibitem{rasiwasia2010new}
Nikhil Rasiwasia, Jose Costa~Pereira, Emanuele Coviello, Gabriel Doyle, Gert~RG Lanckriet, Roger Levy, and Nuno Vasconcelos.
\newblock A new approach to cross-modal multimedia retrieval.
\newblock In {\em Proceedings of the 18th ACM international conference on Multimedia}, pages 251--260, 2010.

\bibitem{ray2023cola}
Arijit Ray, Filip Radenovic, Abhimanyu Dubey, Bryan Plummer, Ranjay Krishna, and Kate Saenko.
\newblock Cola: A benchmark for compositional text-to-image retrieval.
\newblock {\em Advances in Neural Information Processing Systems}, 36:46433--46445, 2023.

\bibitem{rombach2022high}
Robin Rombach, Andreas Blattmann, Dominik Lorenz, Patrick Esser, and Bj{\"o}rn Ommer.
\newblock High-resolution image synthesis with latent diffusion models.
\newblock In {\em Proceedings of the IEEE/CVF conference on computer vision and pattern recognition}, pages 10684--10695, 2022.

\bibitem{rubio2017multi}
Antonio Rubio, LongLong Yu, Edgar Simo-Serra, and Francesc Moreno-Noguer.
\newblock Multi-modal joint embedding for fashion product retrieval.
\newblock In {\em 2017 IEEE International Conference on Image Processing (ICIP)}, pages 400--404. IEEE, 2017.

\bibitem{shahbazi2024coverage}
Nima Shahbazi, Mahdi Erfanian, and Abolfazl Asudeh.
\newblock Coverage-based data-centric approaches for responsible and trustworthy ai.
\newblock {\em IEEE Data Eng. Bull.}, 47(1):3--17, 2024.

\bibitem{singh2022flava}
Amanpreet Singh, Ronghang Hu, Vedanuj Goswami, Guillaume Couairon, Wojciech Galuba, Marcus Rohrbach, and Douwe Kiela.
\newblock Flava: A foundational language and vision alignment model.
\newblock In {\em Proceedings of the IEEE/CVF conference on computer vision and pattern recognition}, pages 15638--15650, 2022.

\bibitem{tanaka2019data}
Fabio Henrique Kiyoiti Dos~Santos Tanaka and Claus Aranha.
\newblock Data augmentation using gans.
\newblock {\em arXiv preprint arXiv:1904.09135}, 2019.

\bibitem{tang2023interacting}
Xu~Tang, Yijing Wang, Jingjing Ma, Xiangrong Zhang, Fang Liu, and Licheng Jiao.
\newblock Interacting-enhancing feature transformer for cross-modal remote-sensing image and text retrieval.
\newblock {\em IEEE Transactions on Geoscience and Remote Sensing}, 61:1--15, 2023.

\bibitem{thrush2022winoground}
Tristan Thrush, Ryan Jiang, Max Bartolo, Amanpreet Singh, Adina Williams, Douwe Kiela, and Candace Ross.
\newblock Winoground: Probing vision and language models for visio-linguistic compositionality.
\newblock In {\em Proceedings of the IEEE/CVF Conference on Computer Vision and Pattern Recognition}, pages 5238--5248, 2022.

\bibitem{torabi2016learning}
Atousa Torabi, Niket Tandon, and Leonid Sigal.
\newblock Learning language-visual embedding for movie understanding with natural-language.
\newblock {\em arXiv preprint arXiv:1609.08124}, 2016.

\bibitem{tremblay2018training}
Jonathan Tremblay, Aayush Prakash, David Acuna, Mark Brophy, Varun Jampani, Cem Anil, Thang To, Eric Cameracci, Shaad Boochoon, and Stan Birchfield.
\newblock Training deep networks with synthetic data: Bridging the reality gap by domain randomization.
\newblock In {\em Proceedings of the IEEE conference on computer vision and pattern recognition workshops}, pages 969--977, 2018.

\bibitem{trummer2022codexdb}
Immanuel Trummer.
\newblock Codexdb: Synthesizing code for query processing from natural language instructions using gpt-3 codex.
\newblock {\em Proceedings of the VLDB Endowment}, 15(11):2921--2928, 2022.

\bibitem{trummer2023demonstrating}
Immanuel Trummer.
\newblock Demonstrating gpt-db: Generating query-specific and customizable code for sql processing with gpt-4.
\newblock {\em Proceedings of the VLDB Endowment}, 16(12):4098--4101, 2023.

\bibitem{villaizan2024diffusion}
Mario Villaiz{\'a}n-Vallelado, Matteo Salvatori, Carlos Segura, and Ioannis Arapakis.
\newblock Diffusion models for tabular data imputation and synthetic data generation.
\newblock {\em arXiv preprint arXiv:2407.02549}, 2024.

\bibitem{wang2015image}
Jian Wang, Yonghao He, Cuicui Kang, Shiming Xiang, and Chunhong Pan.
\newblock Image-text cross-modal retrieval via modality-specific feature learning.
\newblock In {\em Proceedings of the 5th ACM on International Conference on Multimedia Retrieval}, pages 347--354, 2015.

\bibitem{wang2015cluster}
Shuhui Wang, Fuzhen Zhuang, Shuqiang Jiang, Qingming Huang, and Qi~Tian.
\newblock Cluster-sensitive structured correlation analysis for web cross-modal retrieval.
\newblock {\em Neurocomputing}, 168:747--760, 2015.

\bibitem{wang2025cross}
Tianshi Wang, Fengling Li, Lei Zhu, Jingjing Li, Zheng Zhang, and Heng~Tao Shen.
\newblock Cross-modal retrieval: a systematic review of methods and future directions.
\newblock {\em Proceedings of the IEEE}, 2025.

\bibitem{wang2020minilm}
Wenhui Wang, Furu Wei, Li~Dong, Hangbo Bao, Nan Yang, and Ming Zhou.
\newblock Minilm: Deep self-attention distillation for task-agnostic compression of pre-trained transformers.
\newblock {\em Advances in neural information processing systems}, 33:5776--5788, 2020.

\bibitem{woo2023convnext}
Sanghyun Woo, Shoubhik Debnath, Ronghang Hu, Xinlei Chen, Zhuang Liu, In~So Kweon, and Saining Xie.
\newblock Convnext v2: Co-designing and scaling convnets with masked autoencoders.
\newblock In {\em Proceedings of the IEEE/CVF conference on computer vision and pattern recognition}, pages 16133--16142, 2023.

\bibitem{xin2023improving}
Yifei Xin, Dongchao Yang, and Yuexian Zou.
\newblock Improving text-audio retrieval by text-aware attention pooling and prior matrix revised loss.
\newblock In {\em ICASSP 2023-2023 IEEE International Conference on Acoustics, Speech and Signal Processing (ICASSP)}, pages 1--5. IEEE, 2023.

\bibitem{xu2022regnet}
Jing Xu, Yu~Pan, Xinglin Pan, Steven Hoi, Zhang Yi, and Zenglin Xu.
\newblock Regnet: self-regulated network for image classification.
\newblock {\em IEEE Transactions on Neural Networks and Learning Systems}, 34(11):9562--9567, 2022.

\bibitem{yang2021rethinking}
Yang Yang, Chubing Zhang, Yi-Chu Xu, Dianhai Yu, De-Chuan Zhan, and Jian Yang.
\newblock Rethinking label-wise cross-modal retrieval from a semantic sharing perspective.
\newblock In {\em IJCAI}, pages 3300--3306, 2021.

\bibitem{ye2024saliencycut}
Jianan Ye, Yijie Hu, Xi~Yang, Qiu-Feng Wang, Chao Huang, and Kaizhu Huang.
\newblock Saliencycut: Augmenting plausible anomalies for anomaly detection.
\newblock {\em Pattern Recognition}, 153:110508, 2024.

\bibitem{yianilos1993data}
Peter~N Yianilos.
\newblock Data structures and algorithms for nearest neighbor search in general metric spaces.
\newblock In {\em Soda}, volume~93, pages 311--21, 1993.

\bibitem{yu2020bdd100k}
Fisher Yu, Haofeng Chen, Xin Wang, Wenqi Xian, Yingying Chen, Fangchen Liu, Vashisht Madhavan, and Trevor Darrell.
\newblock Bdd100k: A diverse driving dataset for heterogeneous multitask learning.
\newblock In {\em Proceedings of the IEEE/CVF conference on computer vision and pattern recognition}, pages 2636--2645, 2020.

\bibitem{zhang2023text}
Chenshuang Zhang, Chaoning Zhang, Mengchun Zhang, and In~So Kweon.
\newblock Text-to-image diffusion models in generative ai: A survey.
\newblock {\em arXiv preprint arXiv:2303.07909}, 2023.

\bibitem{zhang2010interactive}
Lanbo Zhang and Yi~Zhang.
\newblock Interactive retrieval based on faceted feedback.
\newblock In {\em Proceedings of the 33rd international ACM SIGIR conference on Research and development in information retrieval}, pages 363--370, 2010.

\bibitem{zhang2019editing}
Rui Zhang, Tao Yu, He~Yang Er, Sungrok Shim, Eric Xue, Xi~Victoria Lin, Tianze Shi, Caiming Xiong, Richard Socher, and Dragomir Radev.
\newblock Editing-based sql query generation for cross-domain context-dependent questions.
\newblock {\em arXiv preprint arXiv:1909.00786}, 2019.

\bibitem{zhang2024recognize}
Youcai Zhang, Xinyu Huang, Jinyu Ma, Zhaoyang Li, Zhaochuan Luo, Yanchun Xie, Yuzhuo Qin, Tong Luo, Yaqian Li, Shilong Liu, et~al.
\newblock Recognize anything: A strong image tagging model.
\newblock In {\em Proceedings of the IEEE/CVF Conference on Computer Vision and Pattern Recognition}, pages 1724--1732, 2024.

\bibitem{zhou2003relevance}
Xiang~Sean Zhou and Thomas~S Huang.
\newblock Relevance feedback in image retrieval: A comprehensive review.
\newblock {\em Multimedia systems}, 8:536--544, 2003.

\bibitem{zou2023object}
Zhengxia Zou, Keyan Chen, Zhenwei Shi, Yuhong Guo, and Jieping Ye.
\newblock Object detection in 20 years: A survey.
\newblock {\em Proceedings of the IEEE}, 111(3):257--276, 2023.

\end{thebibliography}

\clearpage
\appendix
\section*{APPENDIX}
\section{Discussions and Limitations}\label{sec:lim}

\stitle{Reliance on Existing Models} %
\system relies on Foundation Models for guide tuple generation and on embedders for generating vector representations of images, which may have their own capabilities, limitations, and inherent biases. This reliance presents both strengths and weaknesses in \system's architecture. The strength lies in its ability to improve performance by upgrading the internal Foundation Models and embedders. On the other hand, though, the limitations of these models also limit \system. To mitigate this issue, \system supports multiple Foundation Models and embedders, enabling it to draw from a broader range of knowledge and reduce the impact of individual model limitations.

\stitle{Preprocessing Efficiency}
During the preprocessing phase, we employ a predefined set of embedders to generate vector representations of images. 
\system processes the images in parallel while utilizing the available GPU resources to distribute the workload efficiently. 
Moreover, \system supports multiple operational modes upon installation. For instance, the fast mode utilizes only two embedders which offers preprocessing speeds that are comparable to those of state-of-the-art methods.

\stitle{Inference Efficiency} 
\system's inference consists of three main sequential operations: (a) generating guide tuples, (b) k-NN lookup, and (c) ranking aggregation.
Theoretically, for retrieving $k$ results from a dataset of size $n$ with constants $m$ (number of guide images) and $\ell$ (number of embedders), Step (a) is a constant-time operation. Step (b)'s time complexity depends on the underlying vector store algorithm, and in the current implementation, we leverage HNSW data structure index \cite{malkov2018efficient}, which provides k-NN lookup time complexity of $O(k\log(n))$,
while Step (c) uses Fagin's instance-optimal TA algorithm for rank aggregation~\cite{fagin2001optimal}.

In practice, however, the bottleneck lies in Step (a), generating guide tuples. For instance, in all of our experiments with up to $n=120K$ images, steps (b) and (c) took less than a second, dominated by the image generation time.
Fortunately, the image generation time is independent of the dataset size $n$ or the output size $k$,  and only relies on the number of guide images $m$. 
Moreover, the generation process can run in parallel, significantly improving performance. \system's current implementation supports both on-device and online image generation models. For optimal performance, we recommend using fast on-device generation models to eliminate network latency. Moreover, as demonstrated in our ablation study, \system maintains high accuracy even with lower-quality, lower-resolution images. Therefore, employing fast, lower-quality foundation models can significantly accelerate the overall process. In addition to these, the builtin Query Complexity Classifier (\ref{sec:prac:complexity}) enables \system to generate images only when the query deemed complex for preliminary methods, which reduces the average inference time drastically compared to using \system for all queries.

\stitle{Extensions to Other Modalities}
The core idea behind \system is adaptable to other modalities beyond images, including audio and video. However, due to the current limitations in publicly available foundation models, our focus in this paper was on the image data. As foundation models for these additional modalities continue to improve and become more accessible, we plan to extend \system to support them. This will allow \system to operate across multiple data types, unlocking further potential for handling complex, multi-modal queries in the near future.

\section{Proof of Theorem~\ref{th:errorbound}}\label{proof:th1}

\noindent{\bf Theorem~\ref{th:errorbound}}
{\it
    Let \(\delta_{\nlq,i} = \delta\left(\vec{q}(\nlq),\vec{v}^o_i\right)\), $\forall t_i\in \dee$.
    Given a positive value $\gamma$,
    \[
    \Pr\left(
    \left(\frac{\bar{\delta}_{\nlq,i}}{\delta_{\nlq,i}}\geq (1+\gamma)\right) \vee 
    \left(\frac{\bar{\delta}_{\nlq,i}}{\delta_{\nlq,i}}\leq (1-\gamma)\right)
    \right) \leq \mathbf{e}^{\frac{-m\,l\,\gamma^2\delta_{\nlq,i}}{3}} + \mathbf{e}^{\frac{-m\,l\,\gamma^2\delta_{\nlq,i}}{2}}
    \]
}

\stitle{Proof} 
Replacing $\bar{\delta}_{\nlq,i}$ with the right-hand side of Equation~\ref{eq:distanceEstimation}, we get
    \begin{align*}
    \Pr\left(\frac{\bar{\delta}_{\nlq,i}}{\delta_{\nlq,i}}\geq (1+\gamma)\right) &= \Pr\left(\frac{\frac{1}{m\, l}\sum_{j=1}^m\sum_{\ell=1}^l ~\delta\left( \embedding^\ell(\bar{\gee}_j), \vec{v}^\ell_i\right)}{\delta_{\nlq,i}}\geq (1+\gamma)\right)\\
    &= \Pr\left(\sum_{j=1}^m\sum_{\ell=1}^l ~\delta\left( \embedding^\ell(\bar{\gee}_j), \vec{v}^\ell_i\right)
    \geq (1+\gamma) m\, l\, \delta_{\nlq,i}\right)
    \end{align*}

    Since \(\ee\left[\delta^\ell_{i,j}\right] = \delta^o_{i,j},~\forall t_i,t_j\) and \(\ee\left[ \embedding^o(\bar{\gee}_\nlq) \right] = \embedding^o(\gee_\nlq)\),
    \[
    \ee\left[ \delta\left( \embedding^\ell(\bar{\gee}_j), \vec{v}^\ell_i\right)\right]
    =
    \delta_{\nlq,i}
    \]
    
    As a result,

    \[
    \ee\left[ \sum_{j=1}^m\sum_{\ell=1}^l ~\delta\left( \embedding^\ell(\bar{\gee}_j), \vec{v}^\ell_i\right)\right]
    =
    m\,l\,\delta_{\nlq,i}
    \]

    Next, applying Chernoff bound~\cite{motwani1996randomized}, we get
    \[
    \Pr\left(\frac{\bar{\delta}_{\nlq,i}}{\delta_{\nlq,i}}\geq (1+\gamma)\right) \leq \mathbf{e}^{\frac{-m\,l\,\gamma^2\delta_{\nlq,i}}{3}}
    \]

    Similarly,

    \[
    \Pr\left(\frac{\bar{\delta}_{\nlq,i}}{\delta_{\nlq,i}}\leq (1-\gamma)\right) \leq \mathbf{e}^{\frac{-m\,l\,\gamma^2\delta_{\nlq,i}}{2}}
    \]

    Therefore, applying the Union bound~\cite{motwani1996randomized}, we get
    \[
    \Pr\left(
    \left(\frac{\bar{\delta}_{\nlq,i}}{\delta_{\nlq,i}}\geq (1+\gamma)\right) \vee 
    \left(\frac{\bar{\delta}_{\nlq,i}}{\delta_{\nlq,i}}\leq (1-\gamma)\right)
    \right) \leq \mathbf{e}^{\frac{-m\,l\,\gamma^2\delta_{\nlq,i}}{3}} + \mathbf{e}^{\frac{-m\,l\,\gamma^2\delta_{\nlq,i}}{2}}
    \]

\section{Practical Optimization}\label{sec:app:prac}

\subsection{Rankings Aggregation and Embedder Trust Mechanism}\label{sec:prac:agg}
\system has been designed with scalability and flexibility in mind, ensuring it is not restricted to any specific embedder. 
Particularly considering the rapid advancements in vector representations, our system is designed to dynamically add, remove, or upgrade embedders.
This will allow our system to enhance its performance as more advanced embedders become available.

\system aggregates the results of the available embedders in real-time to achieve optimal outcomes. Furthermore, \system monitors the performance of embedders for different queries, adjusting its trust in them based on their historical performance. The performance of embedders may vary depending on the type of image input. For instance, one embedder may excel at distinguishing different breeds of dogs but underperform in fruit classification tasks, while another may perform better in the latter. \system addresses this variability by dynamically adjusting the reliability score of embedders for various query topics.

Consider a (predefined) set of topics $T$, and let the natural language query $\nlq$ belong to the topic $t=t(\nlq)\in T$\footnote{The topic $t$ of query $\nlq$ is identified using a topic classification mechanism or is part of the query input.}.
Let $\mathbf{G} = \{\bar{\gee}_1,\cdots, \bar{\gee}_m\}$ be the set of guide images generated for $\nlq$ and consider the embedders $\mathbf{E} = \{\embedding^1, \cdots, \embedding^l\}$. 
For a guide image $\bar{\gee}_j$, let $R^i_j = \mathcal{R}^i(\nlq, \bar{\gee}_j)$ be the sorted list of $k$ nearest images in $\dee$ to the guide image $\bar{\gee}_j$ according to the embedder $\embedding^i$. 

Let $R^i_j$[r] be the $r$-th element of $R^i_j$.

For each topic $t \in T$, we assign a weight $w_i^t \in [0, 1]$ to each embedder $\embedding^i$, representing the reliability of the embedder for answering the queries within topic $t$. 
The weights are normalized so that for each topic $t$,
\[
\sum_{i=1}^{l} w_i^t = 1.
\]

\stitle{Aggregation}
Given the ranked lists from all the embedders and the topic $t$ of the query $\nlq$, the objective is to aggregate these lists into a single final ranking $R_{\nlq}$ by incorporating the topic-specific reliability weights $w^t_i$. 

Let $I_h \in \dee$ be a tuple returned by at least one of the embedders. Let $rank(I_h, R^i_j)$ be the rank assigned to $I_h$ according to the list $R^i_j$. 
Let $\mathsf{S}:[k]\rightarrow[0,1]$ be a monotonically decreasing function that specifies the importance of each rank position in the top-$k$. In particular, while being agnostic to the choice of $S(.)$, following the literature~\cite{cummins2012inference, arampatzis2011modeling, faria2010learning}, we adopt the following position-importance function in \system:
\[
\mathsf{S}(i)= \begin{cases}
                        \frac{1}{i} &~i\leq k \\
                        0 &~\text{otherwise}
                    \end{cases}
\]

Then, the aggregated score of $I_h$ is computed as:

\begin{align}
    score(I_h) = \sum_{i:\embedding^i\in \mathbf{E}} w^t_i \cdot \mathsf{S}(rank(I_h, R^i_j))\label{eq:score}
\end{align}

The top-$k$ tuples based on Equation~\ref{eq:score} are returned as the query output, denoted as $R_{\nlq}$. 

\stitle{Dynamic Reliability-weight Adjustment}
Suppose after showing the top-$k$ results $R_{\nlq}$ to the user, we receive feedback from him/her, which is a subset of results marked as irrelevant to the query $\nlq$ \footnote{We acknowledge that such feedback may not commonly be available after the system deployment. However, crowdsourced feedback on a small set of topic-based benchmarks can be employed to calculate these weights before deploying a new embedder to our system.}.

Let $\mathcal{T} = \{ I'_1, I'_2, \dots, I'_s \}$ be the set of irrelevant tuples where $\mathcal{T} \subseteq R_{\nlq}$

Using this feedback, we update the reliability weights of the embedders for each topic using the Multiplicative Weight Update Method (MWUM) \cite{arora2012multiplicative}. 
Let $t=t(\nlq)$ be the topic of the query $\nlq$.
Then the partial loss of the embedder $\embedding^i$ for a topic $t$ based on $\nlq$ is computed as,

\begin{align}
    \Delta loss_{\nlq}(t,i) = \sum_{j:\bar{\gee}_j \in \mathbf{G}}\sum_{I'_h \in \mathcal{T}}  \mathsf{S}(rank(I'_h, R^i_j))
\end{align}
We only penalize those embedders that highly ranked the irrelevant results within the query topic $t$.

The topic-specific weights are updated multiplicatively:
\[
w_i^t(t+1) = w_i^t(t) \cdot (1 - \eta \cdot \Delta loss_{\nlq}(t,i))
\]
where $\eta \in (0, 1)$ is the learning rate controlling how much the trust in each embedder changes over time.
Finally, the weights are normalized for topic $t$ to ensure $\sum_{i=1}^{l} w_i^t = 1$.

\subsection{Anomaly Detection in Query Images} \label{sec:prac:anomaly}

In addition to rankings aggregation, \system employs an outlier detection mechanism to identify potentially anomalous guide images based on their embeddings from multiple embedders. These anomalies might indicate data quality issues or images that are inherently different from the majority. This process combines the embeddings from all available embedders and uses a weighted approach to adjust for the reliability of each embedder.

We are given a set of $m$ guide images $\mathbf{G} = \{\bar{\gee}_1, \bar{\gee}_2, ..., \bar{\gee}_m\}$ and a set of $l$ embedders $\mathbf{E} = \{\embedding^1, \embedding^2, ..., \embedding^l\}$. For each image $\bar{\gee}_j$, the embedder $\embedding^i$ produces a high-dimensional embedding $\vec{v}^i_j = \embedding^i(\bar{\gee}_j)$. Thus, for each image $\bar{\gee}_j$, we have $l$ corresponding embeddings from the embedders: 
\[
\{ \vec{v}^1_{j}, \vec{v}^2_{j}, \dots, \vec{v}^l_{j} \}.
\]
Our goal is to detect if any guide image $\bar{\gee}_j$ is an anomaly by analyzing the embeddings across all embedders.

\stitle{Dimensionality Reduction using UMAP \cite{mcinnes2018umap}}
Given the high-dimensional nature of the embeddings, we first apply dimensionality reduction to project the embeddings into a lower-dimensional space. For each embedder $\embedding^i$, we reduce the dimensionality of the embeddings $\{ \vec{v}^i_{1}, \vec{v}^i_{2}, \dots, \vec{v}^i_{m} \}$ to a lower dimension $d$ using UMAP (Uniform Manifold Approximation and Projection) \cite{mcinnes2018umap}. 

UMAP constructs a weighted $k$-nearest neighbors graph of the high-dimensional points, optimizes the layout in a low-dimensional space, and minimizes the cross-entropy between the two spaces. The reduced embeddings are given by:
\[
\vec{v'}^i_{j} = \text{UMAP}(\vec{v}^i_{j}) \in \mathbb{R}^d
\]
where $d \ll \text{dim}(\vec{v}^i_{j})$.

\stitle{Outlier Detection using LOF \cite{breunig2000lof} }
After dimensionality reduction, we apply the Local Outlier Factor (LOF)\cite{breunig2000lof} algorithm to identify outliers. LOF assigns an outlier score to each data point based on the density of its neighborhood. The LOF score for image $\bar{\gee}_j$ with respect to embedder $\embedding^i$ is defined as:
\[
\text{LOF}^i(\bar{\gee}_j) = \text{LOF}(\vec{v'}^i_{j})
\]
This score indicates how much of an outlier $\bar{\gee}_j$ is in the context of the embeddings from embedder $\embedding^i$, with higher scores representing a greater likelihood of being an outlier.

\stitle{Aggregation of Outlier Scores}
The final outlier score of an guide image $\bar{\gee}_j$ is computed by aggregating the individual scores, weighted with the reliability weight of each embedder:  

\[
S(\bar{\gee}_j) = \sum_{i=1}^{l} w_i \cdot \text{LOF}^i(\bar{\gee}_j)
\]

\stitle{Anomaly Detection Threshold}
In the end, if the outlier score $S(\bar{\gee}_j)$ of an image $\bar{\gee}_j$ exceeds a given threshold $\tau$, the image is flagged as an anomaly:
\[
S(\bar{\gee}_j) > \tau \quad \Rightarrow \quad \bar{\gee}_j \text{ is an anomaly}
\]

The threshold $\tau$ is a hyper-parameter adjusted according to the sensitivity required by the user. Higher values of $\tau$ promote greater diversity in the retrieved images, potentially increasing recall but risking the inclusion of images that do not adequately represent the query. This may result in lower accuracy and decreased precision. In contrast, lower values of $\tau$ force \system to adhere more closely to the query, leading to higher precision in the results but potentially missing some relevant images.

\subsection{Guide Image Caching and Metadata Generation}\label{sec:prac:metadata}

\begin{figure}
    \centering
    \begin{subfigure}[b]{0.30\linewidth}
        \centering
        \includegraphics[width=\linewidth]{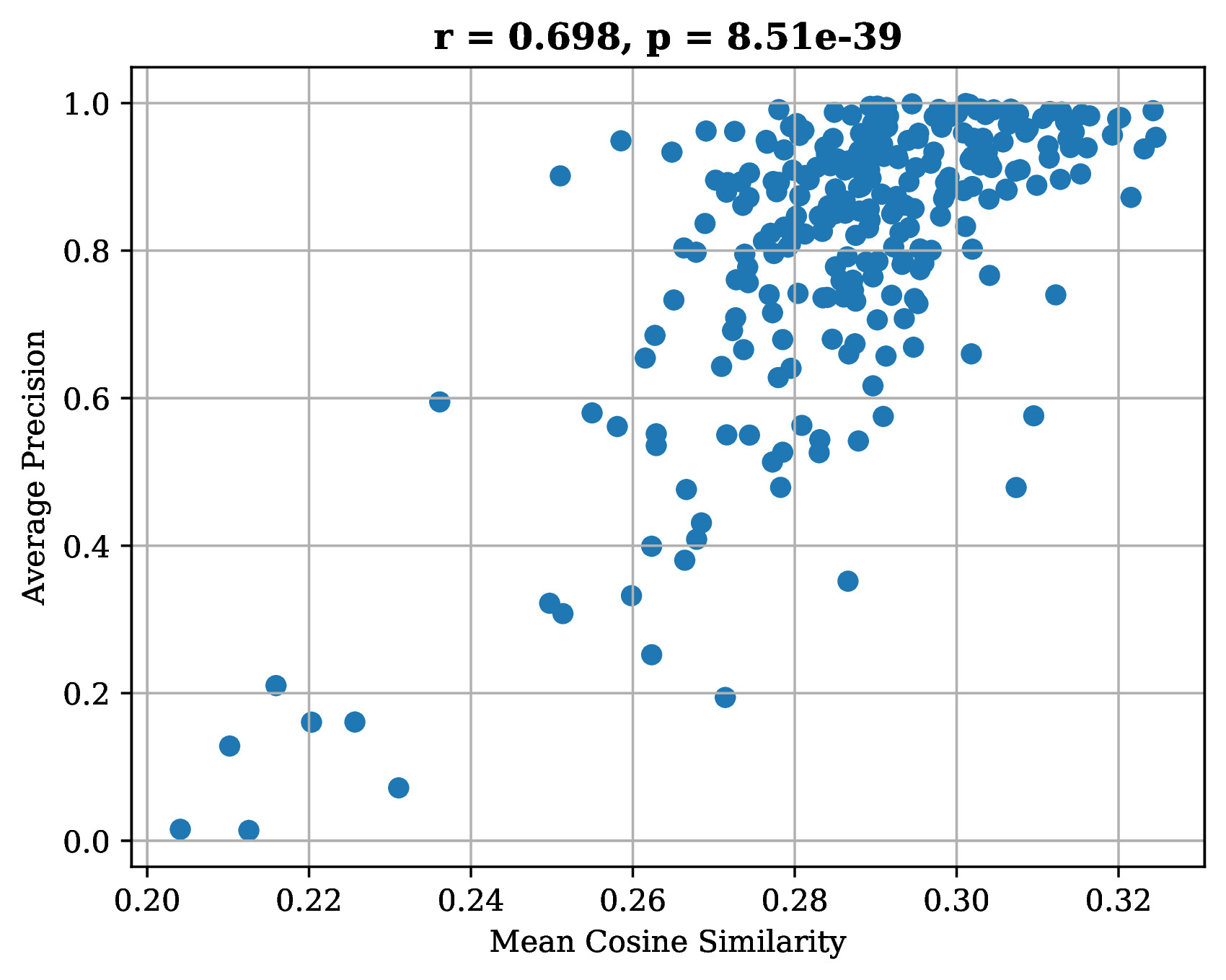}
        \caption{Caltech256}
    \end{subfigure}
    \hfill
    \begin{subfigure}[b]{0.30\linewidth}
        \centering
        \includegraphics[width=\linewidth]{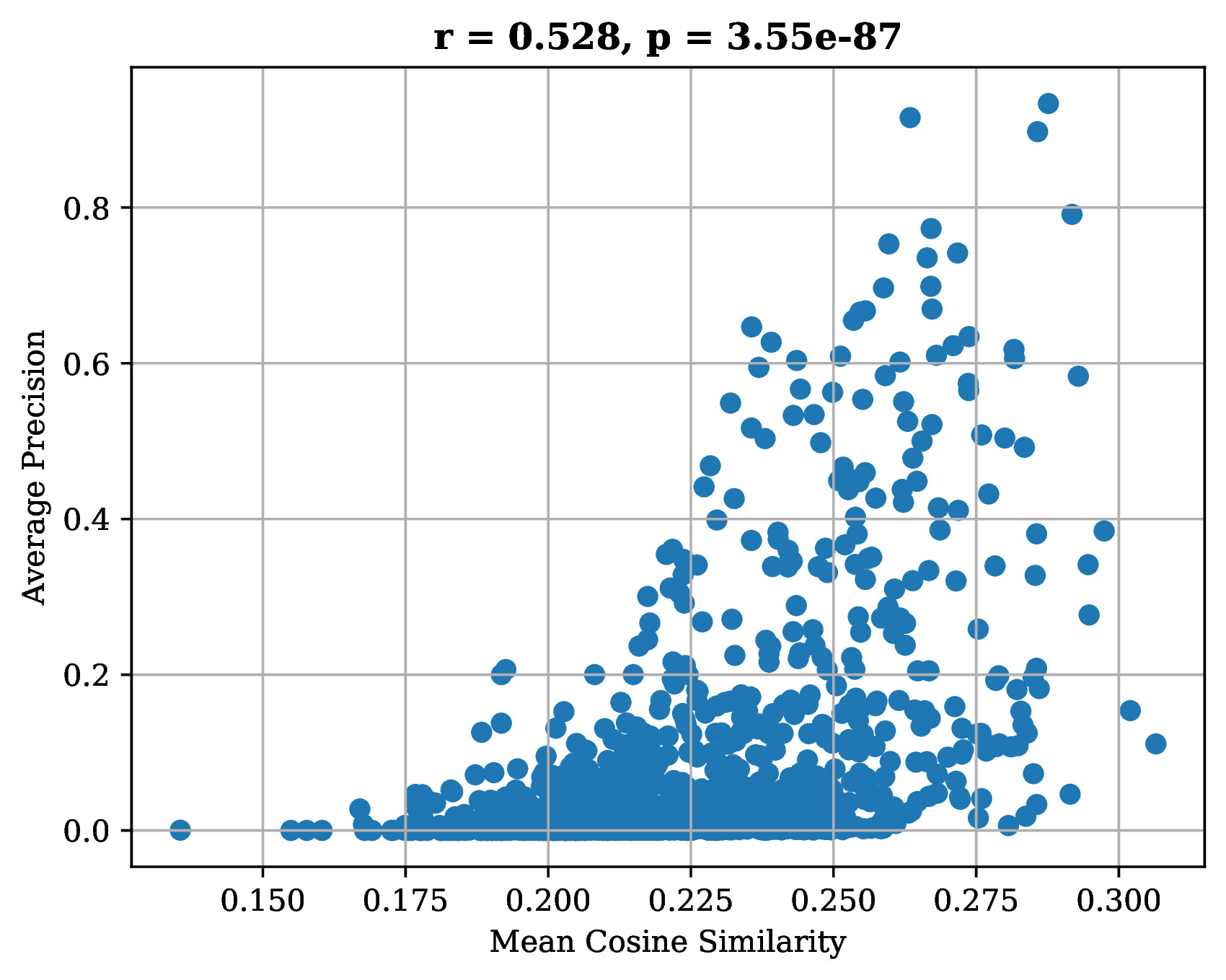}
        \caption{LVIS}
    \end{subfigure}    
    \hfill
    \begin{subfigure}[b]{0.30\linewidth}
        \centering
        \includegraphics[width=\linewidth]{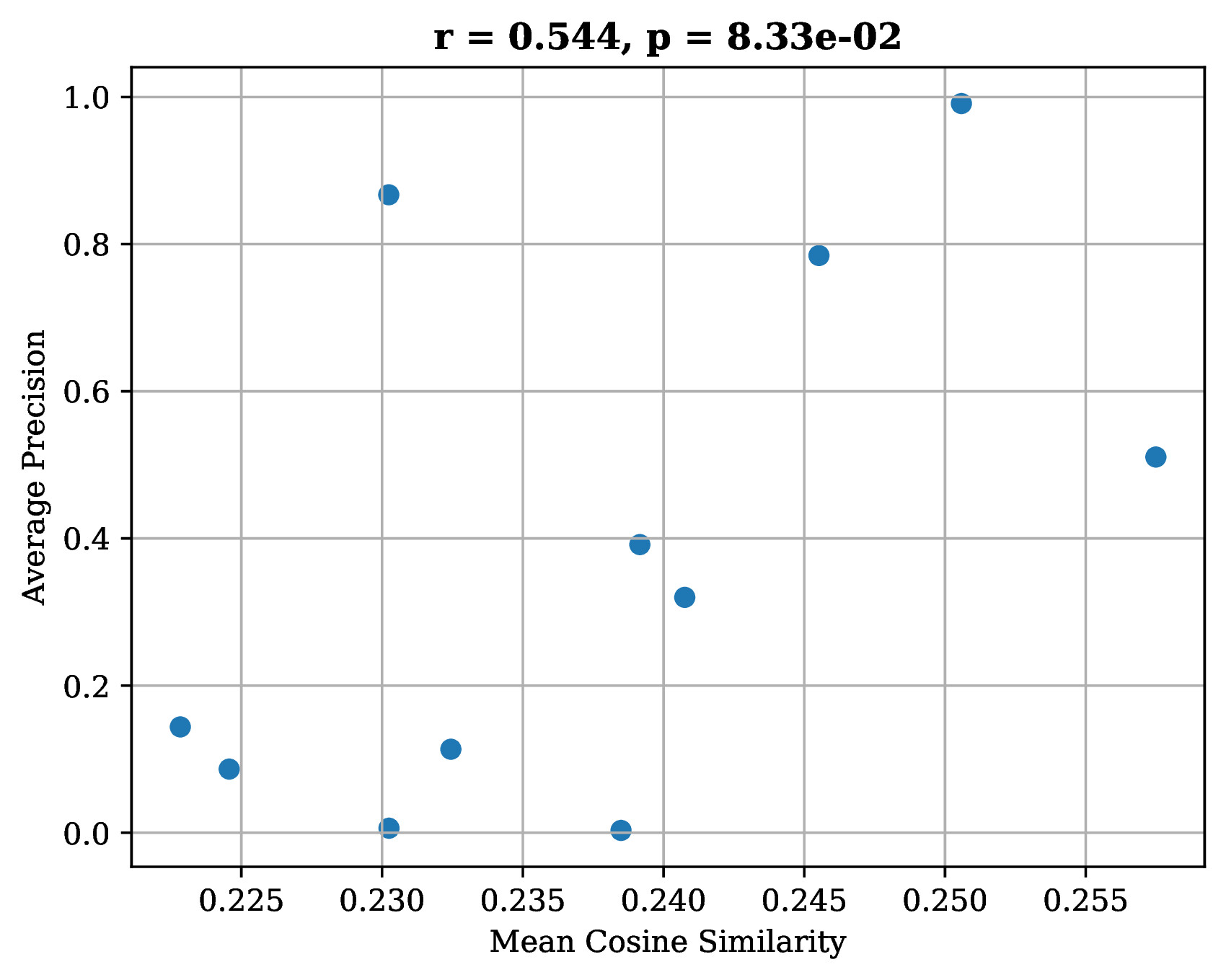}
        \caption{BDD}
    \end{subfigure}    
    \caption{Correlation between CLIP cosine similarity and Mean Average Precision (MAP) across various benchmarks. The Pearson correlation coefficient ($r$) and its associated p-value are displayed within each subplot.}
    \label{fig:cache_clip}
\end{figure}

The generation of guide images, while instrumental for \system's operation, can introduce computational overhead and latency. This is particularly true when utilizing large-scale or proprietary generative models that may incur substantial processing times and network data transmission costs. To mitigate these inefficiencies and enhance overall system performance and cost-effectiveness, \system incorporates a novel caching mechanism designed to minimize the frequency of redundant image generation.

A fundamental challenge in open-domain text-to-image retrieval is the scarcity or inadequacy of rich, comprehensive textual metadata explicitly describing image content. As demonstrated in our experiments (Table~\ref{tab:od-results} and Table~\ref{tab:complex-results}), even advanced image tagging methods often fail to capture the full semantic depth and nuance of visual information. Ideally, the availability of precise metadata (e.g., detailed tags or captions) would enable highly efficient text-based similarity searches to retrieve relevant images directly. However, the manual generation of such high-quality metadata for large image datasets is generally infeasible.

We propose an implicit method for metadata generation and caching by leveraging the search queries themselves. The underlying premise is that for a given input query $\nlq$, any image $I_j$ successfully retrieved by \system for $\nlq$ is highly likely to be semantically related to $\nlq$. This allows us to effectively 'tag' or associate these retrieved images with the input query $\nlq$, thereby enriching our internal knowledge base. This strategy reduces the need for repeated generative processes for similar or semantically related queries, as these tagged images can then be retrieved via a similarity search solely in the text embedding space.

A critical aspect of this query-based metadata generation is quantifying the confidence in the newly established association between a retrieved image and its tagging query, especially given that the granularity of relevant objects or concepts might differ across various queries. To address this, we investigate whether the raw cosine similarity scores produced by a multi-modal embedder, such as CLIP, can serve as a reliable indicator of retrieval confidence. We hypothesize that higher similarity scores for relevant items are predictive of a more confident and accurate retrieval for a given query.

To formally validate this hypothesis, we assess the statistical correlation between an embedder's mean cosine similarity for relevant images and its retrieval performance, as measured by Average Precision (AP). Consider a set of queries $\mathcal{Q} = \{\nlq_k\}_{k=1}^N$. For each query $\nlq_k$, let $\mathcal{R}_{\nlq_k}$ denote the set of ground-truth relevant images. The similarity between vector embeddings of an image $I$ and a query $\nlq_k$ using embedder $\ell$ is computed using cosine similarity, denoted as $ \cos(\angle(\vec{v}_I^\ell, \vec{v}_{\nlq_k}^\ell))$.

We define the mean cosine similarity (MCS) for query $q_k$ from an embedder $\ell$ as:

\begin{align}
    \text{MCS}_\ell(\nlq_k) = \frac{1}{|\mathcal{R}_{\nlq_k}|} \sum_{I \in \mathcal{R}_{\nlq_k}} \cos(\angle(\vec{v}_I^\ell, \vec{v}_{\nlq_k}^\ell))
\end{align}

We then compute the Pearson correlation coefficient, $r$, between the vector of $\{\text{MCS}_\ell(\nlq_k)\}_{k=1}^N$ values and the vector of $\{\text{AP}(\nlq_k)\}_{k=1}^N$ across all queries in $\mathcal{Q}$:

\begin{align}
    r = \frac{\sum_{k=1}^N (\text{MCS}_\ell(\nlq_k) - \overline{\text{MCS}})(\text{AP}(\nlq_k) - \overline{\text{AP}})}{\sqrt{\sum_{k=1}^N (\text{MCS}_\ell(\nlq_k) - \overline{\text{MCS}})^2 \sum_{k=1}^N (\text{AP}(\nlq_k) - \overline{\text{AP}})^2}}
\end{align}

A positive Pearson $r$ value close to +1 indicates a strong linear relationship, suggesting that queries yielding higher mean cosine similarity with their relevant images tend to correspond to higher Average Precision. This correlation validates the utility of embedder similarity scores as a proxy for retrieval confidence. Conversely, a non-significant p-value ($p > 0.05$) would imply no statistically reliable linear relationship in the observed data.

We empirically validate this correlation by calculating $r$ and its corresponding $p$-value for the Caltech256~\cite{griffin_holub_perona_2022}, LVIS~\cite{gupta2019lvis} and BDD100K~\cite{yu2020bdd100k} datasets using CLIP as the embedder. As illustrated in Figure~\ref{fig:cache_clip}, these experiments consistently demonstrate a strong direct correlation between the absolute value of CLIP's cosine similarity and Mean Average Precision (MAP). This observation forms the basis of our image tagging strategy: formally, given an input query $\nlq$ and a fine-tuned threshold $\tau$, \system finds relevant images, and in a post-retrieval stage, assigns query $\nlq$ to all images $I \in \mathcal{R}_{\nlq}$ for which $\cos(\angle (\vec{v}_I^\ell, \vec{v}_\nlq^\ell)) \ge \tau$. We will elaborate on how these generated tags can improve the inference speed of \system in Appendix~\ref{sec:prac:complexity}.

\subsection{Query Complexity Classifier}\label{sec:prac:complexity}
While \system shows outstanding performance and accuracy in retrieval by using guide image generation in its core methodology, we observe that for many simpler queries, a large computational overhead of image generation may not be necessary to achieve satisfactory results. In such cases, computationally efficient retrieval methods, such as conventional contrastive learning approaches like CLIP~\cite{radford2021learning} and ALIGN~\cite{jia2021scaling}, might provide sufficient performance on their own.

To optimize performance and resource utilization, we propose a query complexity classifier designed to dynamically assess the complexity of an input query $\nlq$. This module is designed to predict the expected Average Precision (AP) of a query's retrieval if performed solely using these preliminary methods. These methods, denoted as $\mathcal{M} = \{M_p\}_{p=1}^P$, include established contrastive learning models and our proposed tag-based search mechanism (as introduced in Section~\ref{sec:prac:metadata}), all of which operate without requiring on-the-fly image generation. For an input query $\nlq$, each method $M_p \in \mathcal{M}$ produces a ranked list of images and their associated cosine similarity scores, denoted as $S_{M_p}(I, \nlq) = \cos(\angle(\vec{v}_I, \vec{v}_{\nlq}))$. Specifically, for method $M_p$, the retrieval result can be represented as $R_p(\nlq) = \{(I_{p,j}, S_{M_p}(I_{p,j}, \nlq))\}_{j=1}^N$.

The design of this complexity classifier prioritizes both computational efficiency and contextual relevance. Predicting query complexity solely based on linguistic features of the query text itself can be insufficient, as a query's actual retrieval effectiveness often depends on the specific content and nuances of the underlying image database being searched. Furthermore, for an on-the-fly decision module that must operate upon query arrival, the prediction process should be extremely fast to avoid becoming a bottleneck. To this end, we use a feature-based approach where features are derived directly from the preliminary retrieval results, which are already computed, and are designed to capture not only the query's inherent characteristics but also its interaction with the database. This allows the classifier to operate as a lightweight model, analyzing readily available metrics rather than performing complex, potentially time-consuming, end-to-end linguistic analysis of the query text.

From the top-$K$ retrieved images and their cosine similarity scores for each preliminary method, we extract a comprehensive set of features that collectively indicate the query's complexity. These features form a vector $\vec{f}(\nlq) \in \mathbb{R}^D$ and include:

\begin{itemize}[leftmargin=*]
    \item \textbf{Mean Top-$K$ Cosine Similarity Scores:} For each method $M_p$, we compute the average cosine similarity score among its top-$K$ retrieved images. Let $R_{p,K}(\nlq) = \{I_{p,j}\}_{j=1}^K$ denote the set of top-$K$ images retrieved by method $M_p$. The mean top-$K$ cosine similarity score is:
    \[
    \bar{S}_{M_p,K}(\nlq) = \frac{1}{K} \sum_{I \in R_{p,K}(\nlq)} S_{M_p}(I, \nlq)
    \]
    As illustrated in Figure~\ref{fig:cache_clip}, a higher $\bar{S}_{M_p,K}(\nlq)$ generally suggests a higher expected AP for that query.
    \item \textbf{Top-$K$ Inter-Method Overlap Coefficients:} To quantify the consensus among different preliminary methods, we compute overlap metrics between their top-$K$ retrieved sets. For any pair of distinct methods $(M_p, M_q)$ from $\mathcal{M}$, the Jaccard index is used:
    $$J(R_{p,K}(\nlq), R_{q,K}(\nlq)) = \frac{|R_{p,K}(\nlq) \cap R_{q,K}(\nlq)|}{|R_{p,K}(\nlq) \cup R_{q,K}(\nlq)|}$$
    High overlap indicates strong agreement across methods, typically characteristic of queries that yield higher AP values with preliminary models.
    \item \textbf{Confidence Deviation:} We also use measures of consistency in cosine similarity scores within the top-$K$ results for each method. The standard deviation of the top-$K$ cosine similarity scores, $\sigma_{M_p,K}(\nlq)$, serves as an indicator of confidence deviation:
    $$\sigma_{M_p,K}(\nlq) = \sqrt{\frac{1}{K-1} \sum_{I \in R_{p,K}(\nlq)} (S_{M_p}(I, \nlq) - \bar{S}_{M_p,K}(\nlq))^2}$$
    A lower standard deviation suggests higher confidence and uniformity in cosine similarity scores for a given query, which is often observed for queries achieving higher AP.
\end{itemize}

This constructed feature vector $\vec{f}(\nlq)$ is then used as an input into a pre-trained regression model $\mathcal{C}: \mathbb{R}^D \to [0,1]$. For the purpose if real-time operation upon query arrival, $\mathcal{C}$ is designed as a very simple and lightweight machine learning model (e.g., a linear regression model, or a shallow neural network) that operates efficiently on the derived feature vector $\vec{f}(\nlq)$, rather than performing computationally expensive end-to-end text analysis. The regression model outputs a predicted Average Precision score, $\text{AP}_{\text{pred}}(\nlq)$, for the given query using preliminary methods. The model $\mathcal{C}$ is trained using supervised learning on a dataset of queries, where the ground-truth target for each query is its actual Average Precision (AP) score when retrieved using the preliminary methods.

A predefined threshold $\gamma_{AP} \in [0,1]$ is applied to the predicted Average Precision score $\text{AP}_{\text{pred}}(\nlq)$ to define a dynamic routing strategy:
\begin{itemize}[leftmargin=*]
    \item If $\text{AP}_{\text{pred}}(\nlq) \ge \gamma_{AP}$, the query is categorized as "simple." In this case, \system implements a short-circuiting mechanism, directly returning the early retrieval results obtained from the preliminary ensemble. This bypasses the computationally intensive image generation process, significantly reducing latency and computational cost.
    \item If $\text{AP}_{\text{pred}}(\nlq) < \gamma_{AP}$, the query is categorized as "complex." In this scenario, \system proceeds with its full pipeline, including guide image generation and subsequent Monte Carlo sampling, to perform a more complex retrieval.
\end{itemize}

\section{System Development}\label{sec:system:dev}
\subsection{Development Objectives}\label{sec:system:dev:objectives}
As one of our primary objectives, we aim to develop \system as an open-source deployment-ready application that both researchers and developers can easily adopt. 
Specifically, we considered a set of design objectives, outlined in the following; later, in \S~\ref{sec:system:details}, we describe how our system implementation fulfills these objectives.

    \stitle{Database Functionality and CLI Integration} 
    \system should be easily installable across various operating systems and offer an intuitive command-line interface (CLI) for efficient user interaction.
    
    \stitle{Deployment Readiness and Versatility} \system should be engineered for immediate deployment, operating out-of-the-box on a wide range of system configurations, including resource-limited personal computers.
    
    \stitle{Configurability and Abstract Database Modes} Users should have the freedom to select from multiple operational modes (fast, balanced, accurate) that balance retrieval accuracy and latency according to their specific application needs. Additionally, \system should allow for customization beyond these predefined modes.
    
    \stitle{GPU Acceleration and Scalability} Given the computational intensity of image indexing, \system should automatically leverage available GPU resources and support workload distribution across multiple GPUs to ensure scalable performance.
    
    \stitle{Modular Embedder Architecture} As embedders are central to \system's functionality, the architecture should facilitate easy integration and upgrading of new embedder models, ensuring the system remains at the forefront of technological advancements.
    
    \stitle{Multi-Directory Management} \system should be capable of managing multiple image directories concurrently. It should allow users to add, remove, or toggle directories for search while indexing new directories in the background without interrupting ongoing queries.
    
    \stitle{Robustness and Consistency Maintenance} \system should perform consistency checks upon restart after downtime to synchronize any database changes. It should continuously monitor designated directories for modifications, ensuring the database index remains aligned with the file system.
    
    \stitle{Integration with External Image Generators} To facilitate the generation of guide images, \system should provide connection wrappers for leading proprietary and open-source image generation services (e.g., DALL-E, Google Imagen, Replicate). Furthermore, it should allow users to integrate any image generator that complies with the specified request schema.
    
    \stitle{Flexible Output Formats and Enhanced Usability} Recognizing that \system may serve as a component within larger pipelines, it should support multiple output formats, including JSON, YAML, and human-readable text. Additionally, built-in query previews should be provided to improve the overall user experience.

\begin{figure}[htbp]
    \centering
    \begin{subfigure}[b]{0.48\textwidth}
        \centering
        \includegraphics[width=\linewidth, keepaspectratio]{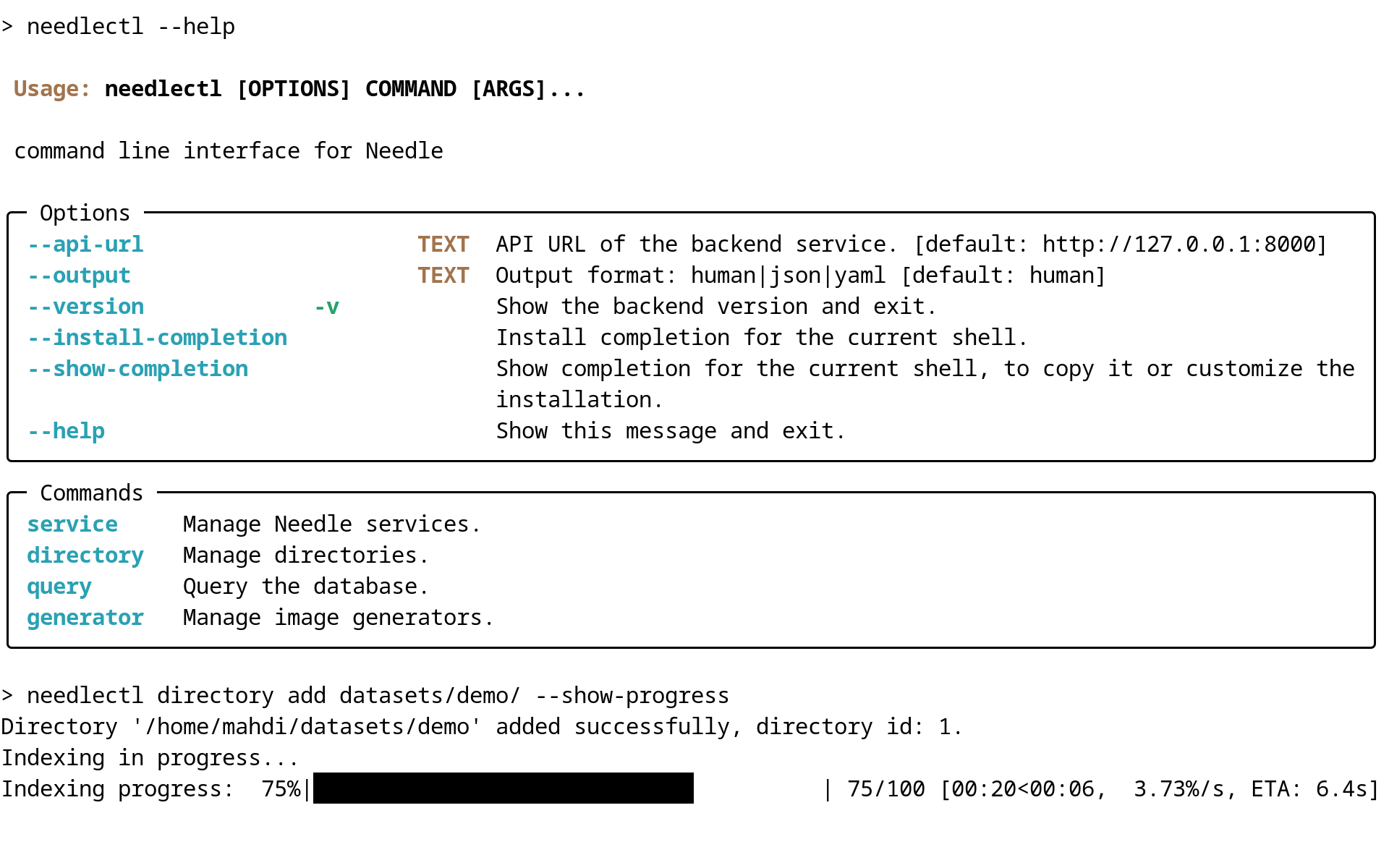}
        \caption{\cli Commands Overview and Adding a Directory}
    \end{subfigure}
    \hfill
    \begin{subfigure}[b]{0.48\textwidth}
        \centering
        \includegraphics[width=\linewidth, keepaspectratio]{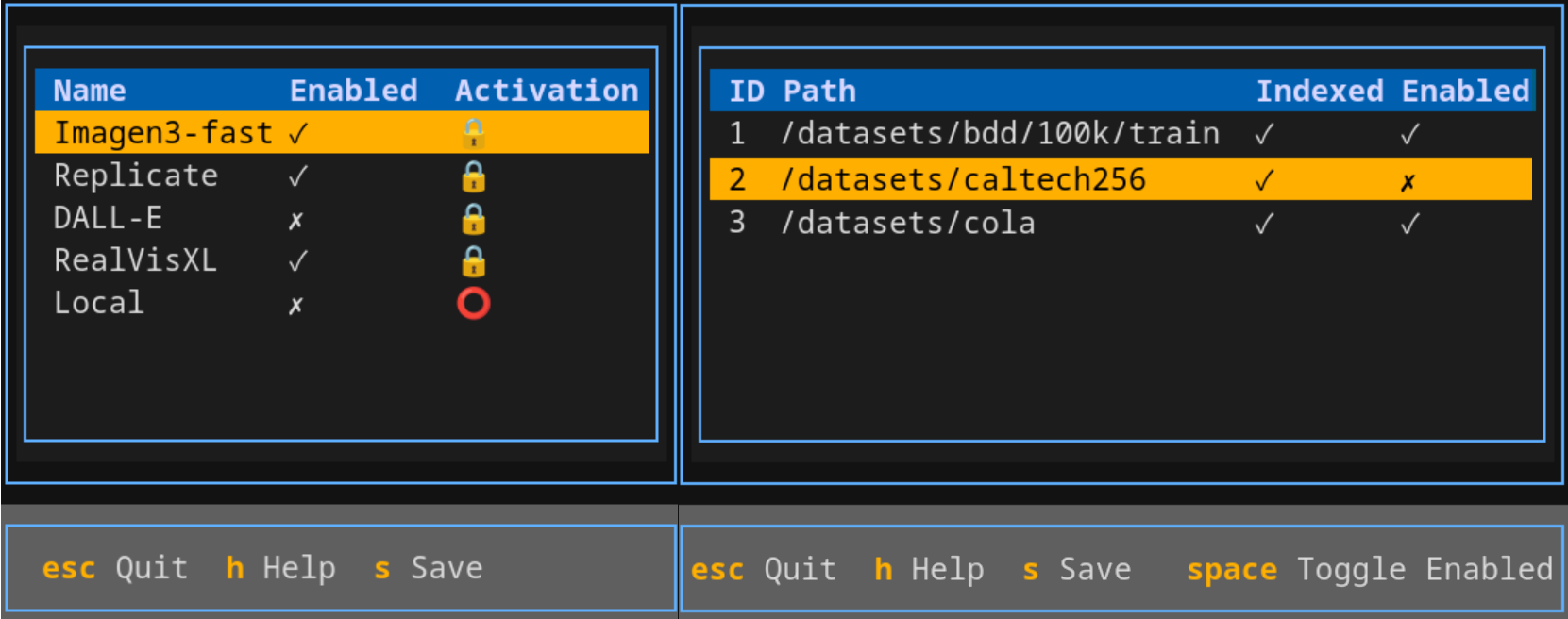}
        \caption{Image Generator/Image Directory Configurations}
    \end{subfigure}
    \vfill
    \begin{subfigure}[b]{0.48\textwidth}
        \centering
        \includegraphics[width=\linewidth, keepaspectratio]{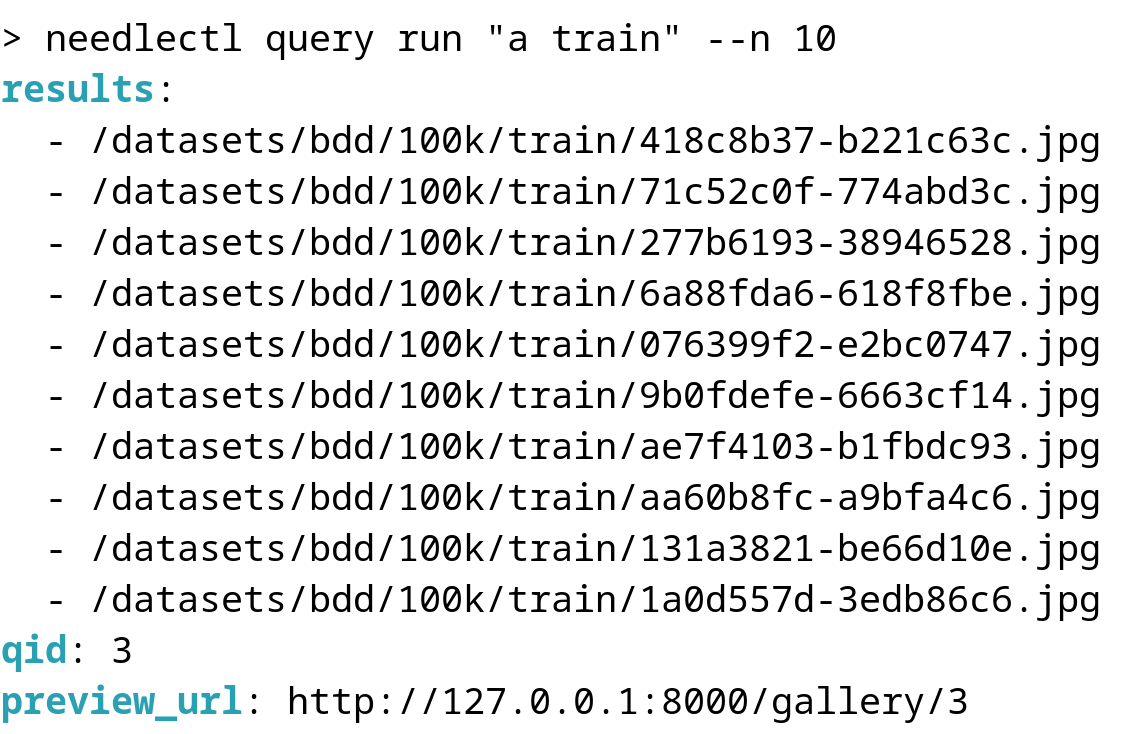}
        \caption{Running queries using \cli and \system Results}
    \end{subfigure}
    \hfill
    \begin{subfigure}[b]{0.48\textwidth}
        \centering
        \includegraphics[width=\linewidth, keepaspectratio]{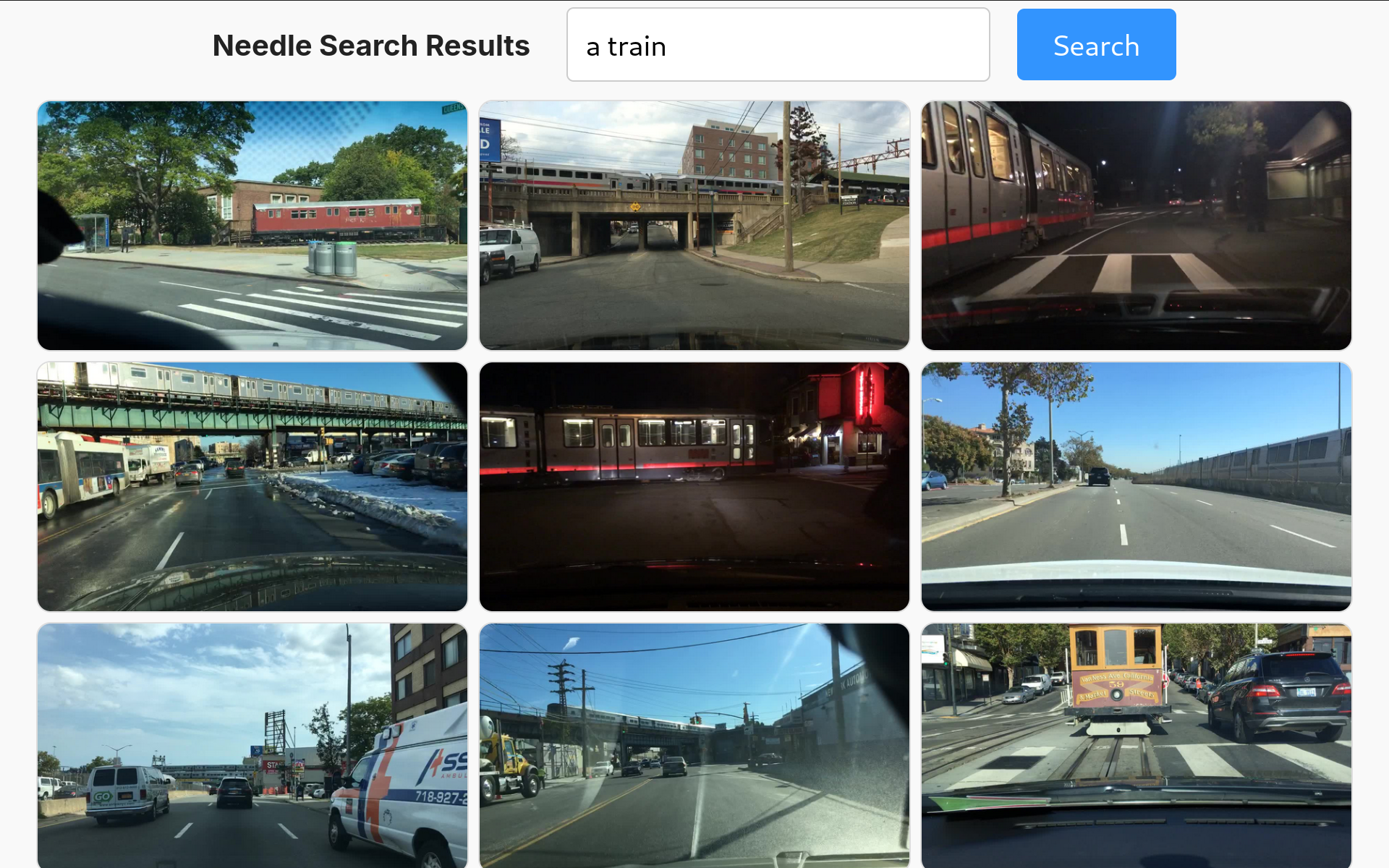}
        \caption{\system Query Results Builtin Preview Page for \hl{\tt[a train]} in BDD dataset~\cite{yu2020bdd100k} -- 102 out of 70K images include a train}
    \end{subfigure}
    \caption{Screenshots of key interfaces and features of \system.}
    \label{fig:screenshots}
\end{figure}

\subsection{System Details}\label{sec:system:details}

Having defined our development objectives, we can now delve into the \system's implementation details.  

\stitle{Architecture} 
\system by nature requires multiple services to work together. It needs a service for handling Image Generation, a service for handling {\it Directory Integration} and {\it Indexing Progress}, and a service for storing {\it calculated embeddings} and {\it kNN lookup}. In order to handle these multiple sources of responsibilities, we have designed \system with a microservice architecture, delegating correlated responsibilities to their respective services. \system's services are: 

\stitle{Backend} This microservice serves as the gateway to the core functionalities of \system. It encapsulates the primary operations of the system, exposing APIs such as \texttt{/search}, \texttt{/directory}, and \texttt{/query}. Additionally, it manages the initialization of embedders and oversees the image indexing process.

\stitle{Database} The backend container needs persistent storage to store directory structures, indexing progress, and other critical metadata. \system is agnostic to the choice of the database, but in the current implementation, we use \texttt{PostgreSQL}\footnote{\url{https://www.postgresql.org/}} for this goal. 

\stitle{Vector Store} \system requires a vector store that supports embedding, indexing, and inner product searches. This service is used to store the outputs of various embedders and to efficiently execute k-nearest neighbor (kNN) searches for a given query. In the current implementation, we have used \texttt{Milvus}\footnote{\url{https://milvus.io/}} for this purpose. 

\stitle{Image Generator Hub} This service manages the image generation process by providing wrappers for renowned services such as DALL-E, Imagen, and Replicate. It also supports custom image generation models that conform to its defined schema. Moreover, the hub offers fallback mechanisms and priority settings to maximize robustness and ensure reliable image generation.

\stitle{Command-Line Interface} \system requires an intuitive interface to work with backend service and apply user's requests. This service is the exposed UI of \system to the end user.

Since the individual services have their own dependencies and requirements, we containerized all core services using \texttt{docker} to mitigate dependency conflicts and make deployment easier across various operating systems. 
We employ \texttt{docker compose} to orchestrate these containers, establish a shared network, and configure the individual services. Additionally, to eliminate the need for direct Docker manipulation, we developed a Command-Line Interface (CLI), \cli, as a standalone binary. \cli installs into the user's path and provides seamless access to the \system core.

\stitle{Technical Stack and Frameworks}
Backend and ImageGeneratorHub services built using \texttt{FastAPI} framework\footnote{\url{https://fastapi.tiangolo.com/}}. For GPU-intensive tasks, \texttt{PyTorch}\footnote{\url{https://pytorch.org/}} is leveraged to distribute workloads across available GPUs, ensuring efficient processing.

Embedder management is streamlined via integrations with Hugging Face and the \texttt{timm} library\footnote{\url{https://github.com/huggingface/pytorch-image-models}}, facilitating the initiation and upgrading of embedding models. The CLI client (\cli) is developed using the \texttt{Typer}\footnote{\url{https://typer.tiangolo.com/}} library to build intuitive command-line applications, while \texttt{textual}\footnote{\url{https://textual.textualize.io/}} is utilized to manage terminal user interfaces (TUIs).

Standalone Python libraries are packaged with \texttt{PyInstaller}\footnote{\url{https://pyinstaller.org/en/stable/}}, simplifying the distribution and execution. Installation, upgrading, and uninstallation processes are automated using \texttt{Bash} scripts, ensuring a smooth user experience. Containerization and orchestration are achieved with \texttt{Docker}\footnote{\url{https://www.docker.com/}} and \texttt{Docker Compose}, respectively.

For documentation, we employ \texttt{mdBook}\footnote{\url{https://rust-lang.github.io/mdBook/}} to create comprehensive and well-structured guides. Finally, all of the build, development, and versioning processes are automated via CI/CD pipelines using GitHub Actions\footnote{\url{https://github.com/features/actions}}.

\stitle{Command-Line Interface and APIs}
\system offers a robust CLI that enables seamless interaction with its core services. The \cli is provided as a standalone binary, allowing users to execute commands directly without requiring additional dependency installations. This design ensures that non-expert users can easily operate \system.
The CLI commands and options have been carefully designed to be intuitive and straightforward. The command structure adheres to the following format:

\begin{verbatim}
needlectl [--global options] [component] [action] [--action options]
\end{verbatim}

This structure supports clear and consistent command usage with the following key components %
(some illustrative examples are provided in Listing~\ref{listing:1}):
\begin{itemize}[leftmargin=*]
    \item \texttt{service}: for managing \system-related tasks
    \item \texttt{directory}: for handling image directory operations
    \item \texttt{query}: for managing active queries %
    \item \texttt{generator}: for interfacing with image-generation functionalities
\end{itemize}

\begin{verbatim}
needlectl service start     # Starts the Needle services
needlectl service log       # Retrieves service logs
needlectl directory add /path/to/image/dir --show-progress
# Adds an image directory for indexing and monitoring, displaying progress
needlectl --output json query run "a wolf" --num-engines 2 --num-images-to-generate 4 --image-quality LOW
# Executes a query using two generators, each generating four low-quality images, with JSON output
\end{verbatim}

\stitle{Installation and Deployment}

\system is installed using a single one-liner command that executes a Bash script. This script automatically verifies system prerequisites and checks for GPU availability. It detects the current operating system and downloads the appropriate configuration. During installation, users are prompted to select a preferred database mode—Fast, Balanced, or Accurate—which optimizes various parameters such as the number of embedders, default image resolution for generation, default number of generators, number of images generated per query, and HNSW\cite{malkov2018efficient} index construction parameters (e.g., \texttt{M} and \texttt{ef}). Finally, the installation script guides users through the initial steps to start and utilize the \system service.

\begin{table}[!tb]
    \centering
    \caption{Embedder Models and Their Weights}
    \label{tab:embedder_weights}
    \begin{tabular}{lcc}
        \toprule
        \textbf{Architecture} & \textbf{Model Name} & \textbf{Weight} \\
        \midrule
        eva\cite{fang2023eva} & {\tt eva02\_large\_patch14\_448}   & 0.8497 \\
        regnet\cite{xu2022regnet} & {\tt regnety\_1280}  & 0.8235 \\
        dinoV2\cite{oquab2023dinov2} & {\tt vit\_large\_patch14\_reg4\_dinov2}  & 0.8235 \\
        CLIP\cite{radford2021learning} & {\tt vit\_large\_patch14\_clip\_33}6   & 0.8146 \\
        convnextV2\cite{woo2023convnext} & {\tt convnextv2\_large} & 0.8184 \\
        bevit\cite{bao2021beit} & beitv2\_large\_patch16\_224 & 0.7660 \\
        \bottomrule
    \end{tabular}
\end{table}

\section{Additional Experiments}

\subsection{Details of Baseline Models} 
\label{sec:appendix_models} 

\stitle{1. CLIP~\cite{radford2021learning}} Developed by OpenAI\footnote{\url{https://huggingface.co/openai/clip-vit-base-patch32}}, this model uses a ViT-B/32 image encoder and a Transformer-based text encoder. It learns a shared embedding space via contrastive learning on large-scale image--text pairs, and its widespread use and benchmarking make it a standard reference.
We use CLIP as our standard baseline and designate categories where CLIP's average precision (AP) is below 0.5 as \emph{"hard categories"}---instances where CLIP struggles to produce satisfactory results. This filtered subset is then used to demonstrate the enhanced capability of alternative baselines in effectively handling rare or difficult queries.
\vspace{1em} %

\stitle{2. ALIGN~\cite{jia2021scaling}} Originally developed by Google, ALIGN (A Large-scale ImaGe and Noisy-text Embedding) employs a dual-encoder architecture similar to CLIP, using separate encoders for images and text trained with a contrastive loss. However, while CLIP (in our \texttt{clip-vit-base-patch32} variant) uses a Vision Transformer for image encoding, ALIGN typically uses a CNN image encoder along with a Transformer for text. Moreover, the original ALIGN was trained on a much larger, noisier dataset (over 1.8 billion image--text pairs) than CLIP. In our experiments, we use an open-source version\footnote{\url{https://huggingface.co/kakaobrain/align-base}} that preserves the core architecture but is trained on different data, achieving even better accuracy on some benchmarks.
\vspace{1em} %

\stitle{3. FLAVA~\cite{singh2022flava}}\footnote{\url{https://huggingface.co/facebook/flava-full}} a unified multimodal model developed by Facebook that jointly learns representations for images and text. Unlike CLIP, which primarily relies on a contrastive approach applied to image-text pairs, and ALIGN, which emphasizes scaling up representation learning using noisy text supervision with separate encoders, FLAVA adopts a holistic pre-training strategy. It integrates both unimodal and multimodal objectives, enabling FLAVA to capture richer semantic interactions across modalities.
\vspace{1em} %

\stitle{4. BLIP + MiniLM} This pipeline approach first converts images into descriptive captions using BLIP~\cite{li2022blip}\footnote{\url{https://huggingface.co/Salesforce/blip-image-captioning-base}}---a state-of-the-art image captioning model from Salesforce known for its high-quality, informative captions. The generated captions are then transformed into embeddings with MiniLM~\cite{wang2020minilm}\footnote{\url{https://huggingface.co/sentence-transformers/all-MiniLM-L6-v2}}, a robust text encoder widely adopted in industrial applications. This decoupled strategy not only leverages mature text retrieval systems but also enables independent optimization of the captioning and text-embedding stages. As a result, it has become a strong baseline and is employed in many commercial products, offering enhanced interpretability and scalability compared to end-to-end models like CLIP.

\subsection{Dataset Details} 
\label{sec:appendix_datasets}

\begin{figure}
    \centering
    \begin{subfigure}[b]{0.49\linewidth}
        \centering
        \includegraphics[width=\linewidth]{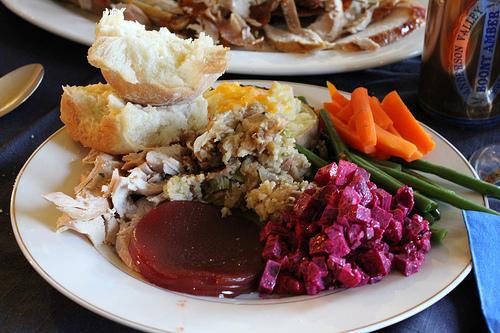}
        \caption{Blue napkin to the right of white plate}
    \end{subfigure}
    \hfill
    \begin{subfigure}[b]{0.49\linewidth}
        \centering
        \includegraphics[width=\linewidth]{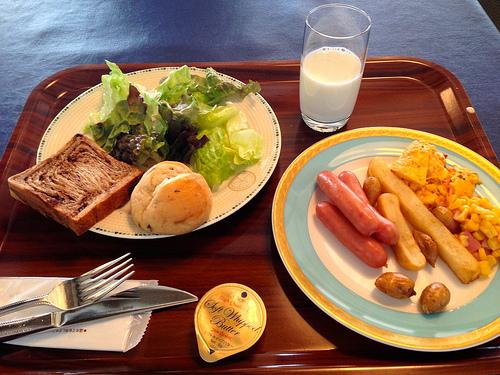}
        \caption{White napkin to the left of blue plate}
    \end{subfigure}    
    \caption{Illustration of COLA compositional benchmark queries, using each image description, where the goal is to identify the correct image}
    \label{fig:cola-example}
\end{figure}

\stitle{Object Detection Datasets}
\begin{enumerate}[leftmargin=*] 
\item {\bf Caltech256~\cite{griffin_holub_perona_2022}}: Contains 30,607 images spanning 256 object categories. Its diverse object classes make it a useful benchmark for assessing the robustness of detection and retrieval models.
\item {\bf MS COCO~\cite{lin2014microsoft}}: With over 118K images and 80 object categories, MS COCO is widely used for object detection and segmentation. Its comprehensive annotations and varied scene compositions provide a challenging testbed.
\item {\bf LVIS~\cite{gupta2019lvis}}: offers instance segmentation with a long-tail distribution of over 1,200 categories. Its focus on rare and fine-grained objects is critical for evaluating retrieval performance on less frequent classes.
\item {\bf BDD100k~\cite{yu2020bdd100k}}: This dataset is tailored to urban driving scenarios with detailed annotations, making it valuable for testing retrieval in real-world, dynamic contexts.
\end{enumerate}

\stitle{Natural Language Query Datasets}

For evaluating the performance of various baselines on complex natural language queries, we focus on scenarios where models must differentiate between two highly similar images that exhibit nuanced differences. In these settings, the model is expected to identify the correct image by carefully attending to subtle details in both the visual content and the corresponding query. To rigorously assess these capabilities, we utilize several compositional benchmarks:
\begin{enumerate}[resume,leftmargin=*] 
\item {\bf COLA~\cite{ray2023cola}}: As a benchmark for compositional text-to-image retrieval, this dataset features nuanced captions that require the system to distinguish between multiple similar images by capturing subtle semantic and spatial details.
\item {\bf Winoground~\cite{thrush2022winoground}}: Designed to evaluate visio-linguistic alignment, Winoground presents pairs of images and captions that differ only in their compositional structure. This dataset challenges models to accurately map nuanced language to the corresponding image, serving as a stringent test of fine-grained retrieval performance.
\item {\bf NoCaps~\cite{agrawal2019nocaps}}: A large-scale benchmark for novel object captioning, featuring images from Open Images with human-generated captions. It is crucial for evaluating a model's ability to handle zero-shot retrieval on images containing objects not commonly found in standard captioning datasets.
\item {\bf SentiCap~\cite{mathews2016senticap}}: Derived from datasets like MS COCO, SentiCap provides images with captions specifically annotated for positive or negative sentiment. It allows for evaluating retrieval systems' ability to understand and retrieve images based on subjective or emotional language cues, moving beyond purely factual descriptions.
\end{enumerate}

\subsection{Evaluation Metrics} 
\label{sec:appendix_metrics} 
We use different evaluation metrics for object detection and complex natural language queries benchmarks.
In object detection benchmarks, we define each benchmark dataset $\mathcal{D}$ as a set of images $\mathcal{D} = \{t_1, t_2, \ldots, t_p\}$, where each image $t_i$ comprises a set of objects $t_i = \{o_1, o_2, \ldots, o_s\}$, with $o_j$ representing an individual object within image $t_i$. For each benchmark, we compute the union of all objects present across all images, denoted as $\bigcup_{i=1}^{n} t_i = \bigcup_{i=1}^{n} \{o_1, o_2, \ldots, o_s\}$, and utilize each unique object $o_j$ within this union as a query $q$ for each retrieval engine. For a given query $q$, we retrieve a ranked list of files $\mathcal{F} = \{f_1, f_2, \ldots, f_n\}$ ($n=60$), representing the retrieved images. We construct a list of relevance scores $\mathcal{R} = \{r_1, r_2, \ldots, r_n\}$, where $r_i = 1$ if $q$ is present in image $f_i$, and $r_i = 0$ otherwise. We start list indexing from 1, so by definition $\mathcal{R}[j] = r_j$. We assume access to a function $C(.)$, which gets an object as input and returns the number of images that this object appears in. We are interested in retrieving at most 10 positive instances. Therefore, we define the effective number of positives as: $ep = \min(10, C(q))$.

Given $\mathcal{R}$ and $ep$, We assess the performance of the retrieval using the following metrics:

\stitle{Mean Recall at $k$ (R@k)} \(
r@k = \frac{\sum_{i=1}^k \mathcal{R}[i]}{ep}
\).
A higher r@k indicates that a larger proportion of the relevant images (up to 10) are retrieved in the top $k$ positions, we report mean r@k (R@k) which is the average of r@k over all objects in the dataset.

\stitle{Precision at $k$ (P@k)} \(
p@k = \frac{\sum_{i=1}^k \mathcal{R}[i]}{k}
\).
A higher p@k reflects that a greater proportion of the retrieved images are relevant, which is crucial when the user examines only the top results, same as before, we report the averaged out P@k over all object in the dataset.

\stitle{Mean Average Precision (MAP)} measures the average precision (AP) over all queries.
\[
AP = \frac{1}{ep} \sum_{i=1}^{n} \left(\frac{\sum_{j=1}^{i} \mathcal{R}[j]}{i}\right) \cdot r_i,
\]
where $r_i = \mathcal{R}[i]$ and the effective number of positives is defined as $ep = \min(10, C(q))$. The overall mean average precision (MAP) is then the average of $AP$ over all objects in the dataset. MAP provides a single-figure measure that considers both the precision and recall across the entire ranked list, rewarding systems that rank relevant images higher.

\stitle{Mean Reciprocal Rank (MRR)} computes the reciprocal rank for each query and then averages these values over all queries. For a given query $q$, let $k_q$ be the smallest index such that $\mathcal{R}[k_q] = 1$, indicating the rank of the first relevant image. The reciprocal rank for $q$ is defined as:
\[
RR(q) =
\begin{cases}
\frac{1}{k_q}, & \text{if } \exists \, k \text{ with } \mathcal{R}[k] = 1,\\[5pt]
0, & \text{otherwise.}
\end{cases}
\]
The Mean Reciprocal Rank (MRR) is the average value over all queries. MRR emphasizes the importance of retrieving at least one relevant image as early as possible for ranking.

For the complex natural language query experiments, we evaluate our method using both MRR and Pairing Accuracy. In these experiments, the baseline is given two images, $t_1$ and $t_2$, along with two captions, $c_1$ and $c_2$, and each caption must be assigned to its corresponding image correctly. Pairing Accuracy is defined as the number of correct assignments divided by the total number of queries. Note that the pairing accuracy for a random baseline is 0.25\% (chance of choosing the correct combination over all possible combinations).

\subsection{Hyper-parameter Ablation Study}\label{sec:exp:ablation}

\begin{figure*}[htbp]
    \centering
    \begin{subfigure}[b]{0.49\linewidth}
        \centering
        \includegraphics[width=\linewidth]{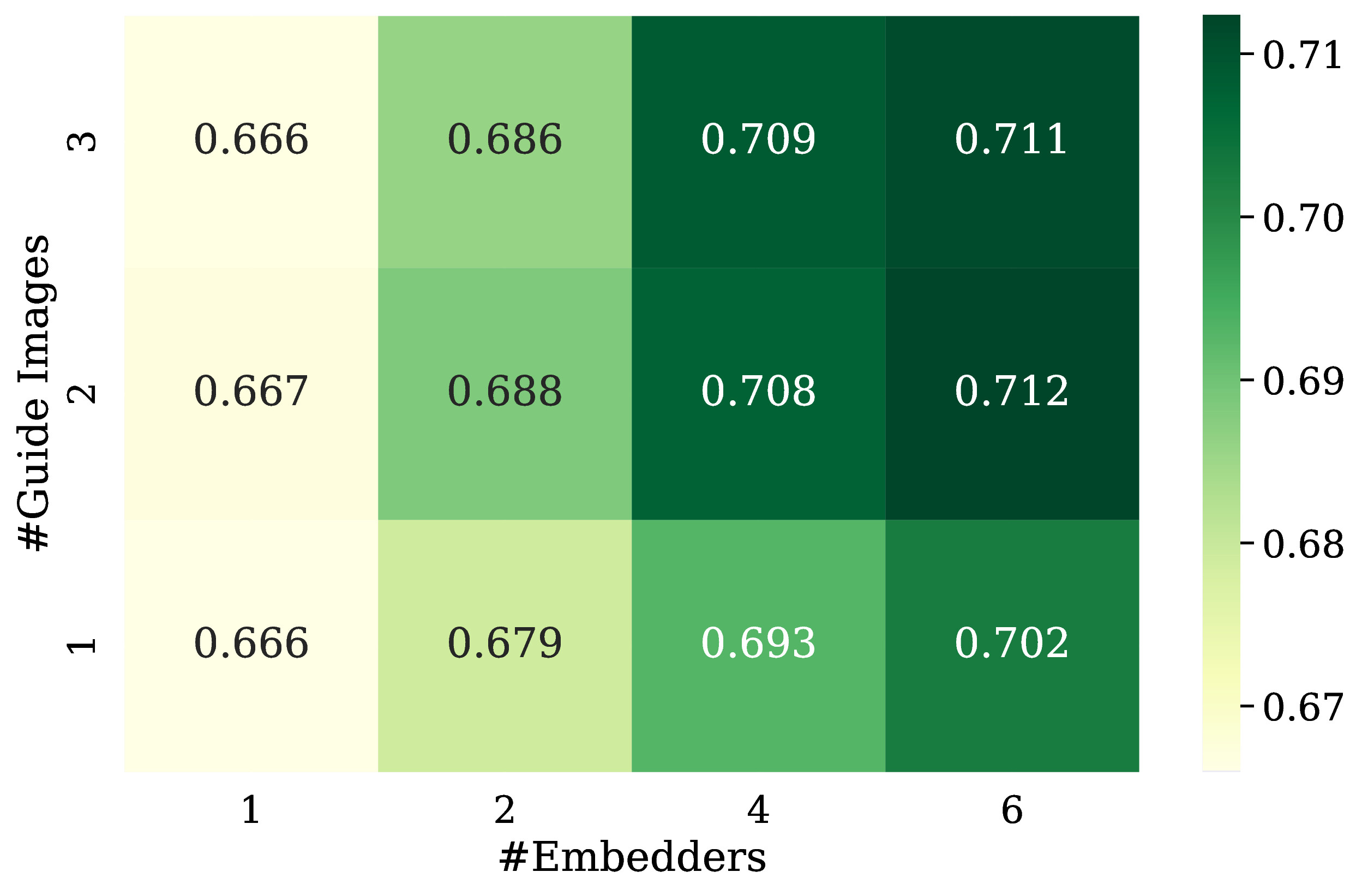}
        \caption{\small BDD}
    \end{subfigure}
    \hfill
    \begin{subfigure}[b]{0.49\linewidth}
        \centering
        \includegraphics[width=\linewidth]{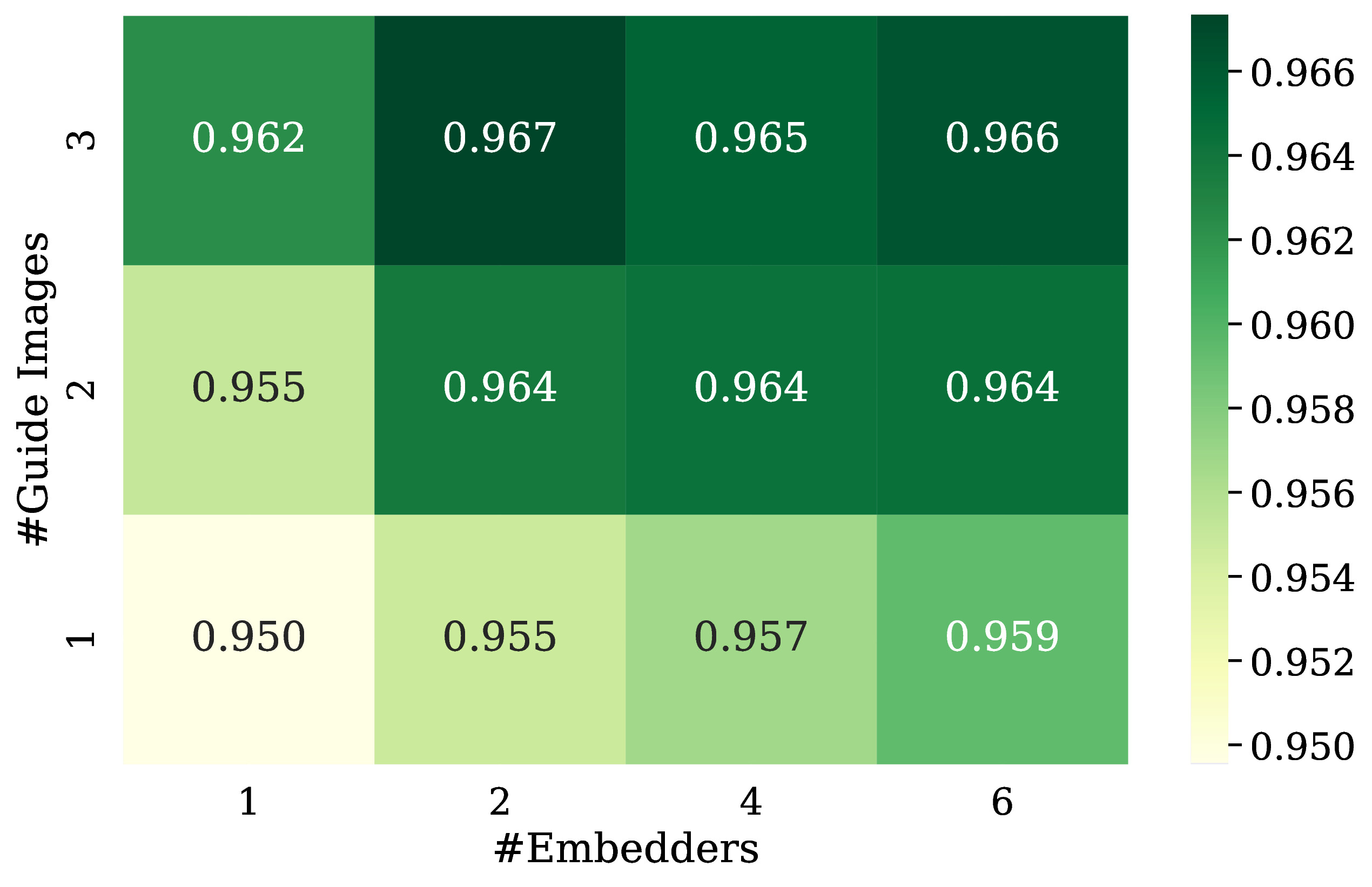}
        \caption{\small Caltech256}
    \end{subfigure}
    \vfill
    \begin{subfigure}[b]{0.49\linewidth}
        \centering
        \includegraphics[width=\linewidth]{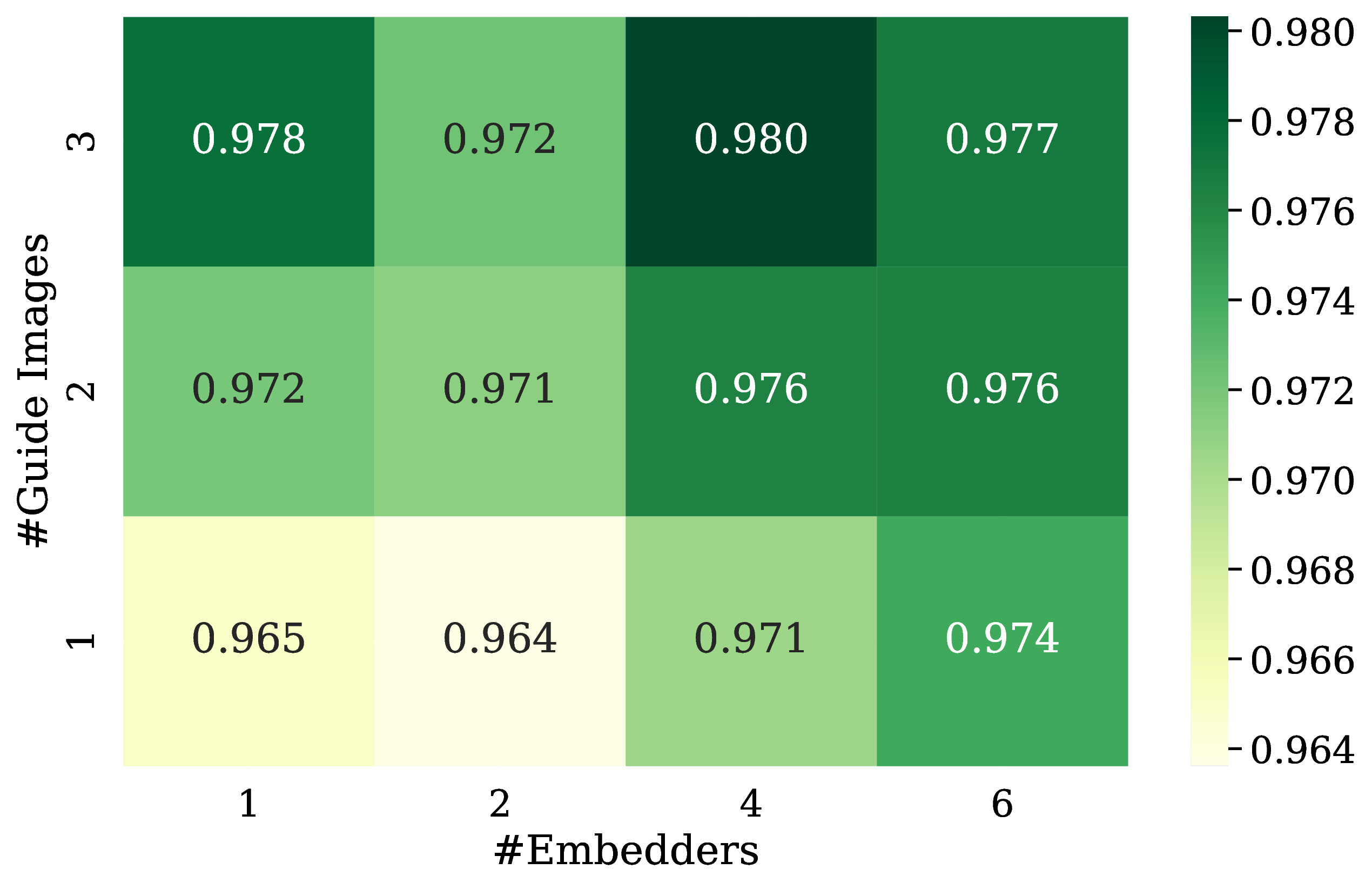}
        \caption{\small COCO}
    \end{subfigure}
    \hfill
    \begin{subfigure}[b]{0.49\linewidth}
        \centering
        \includegraphics[width=\linewidth]{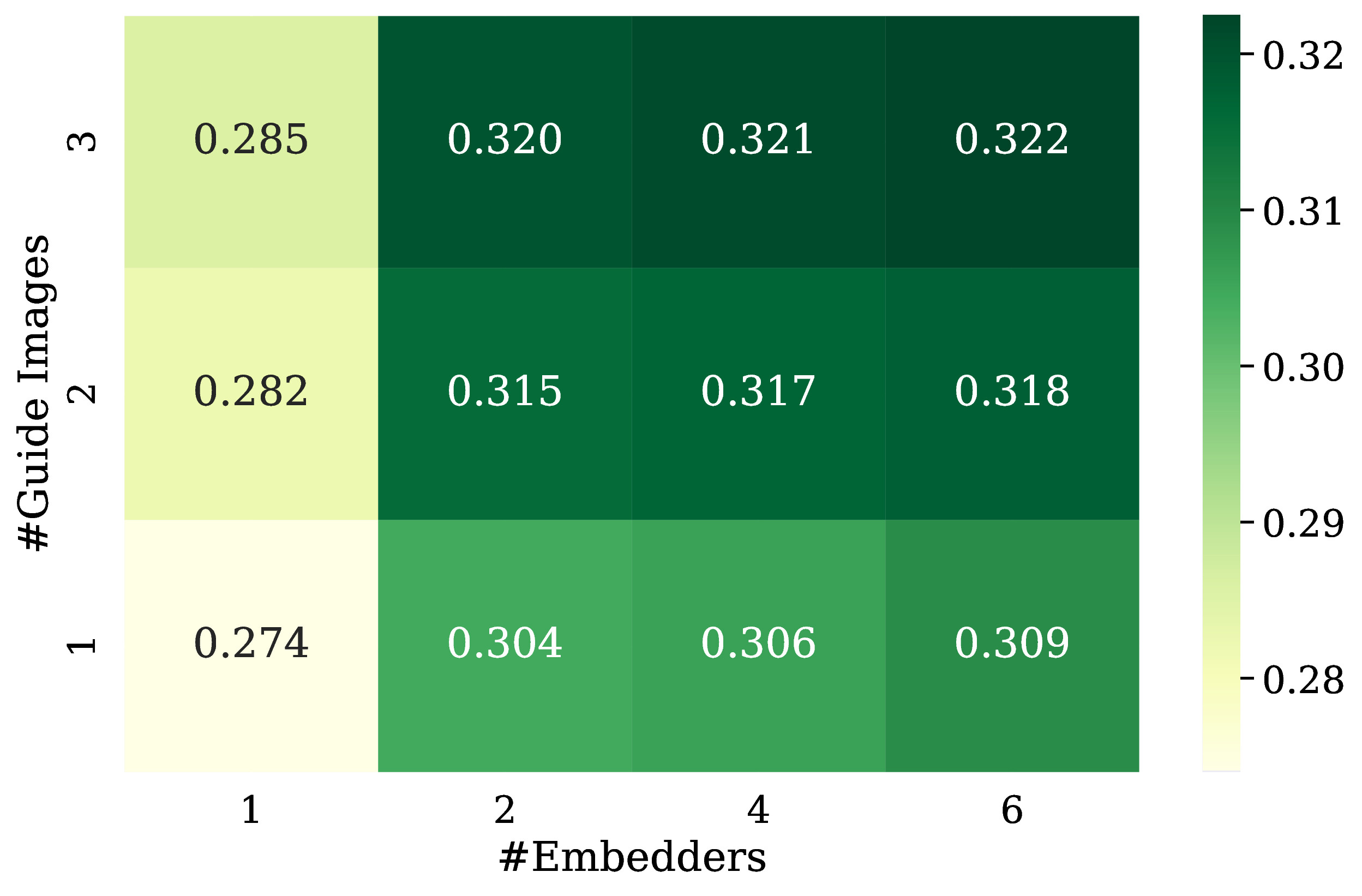}
        \caption{\small LVIS}
    \end{subfigure}
    \caption{Illustration of how varying number of guide images and number of embedders involved affects \system MAP }
    \label{fig:ablation-m-l}
\end{figure*}

In this section, we examine the impact of each hyper-parameter on \system's performance. 

\stitle{Generated Guide Images and Embedders Count}
We investigate the impact of the number of images generated per engine ($m$) and the number of embedders used ($\ell$) on the performance of \system. We vary $m$ over $\{1, 2, 3\}$ and $\ell$ over $\{1, 2, 4, 6\}$, and conduct experiments with these configurations on the BDD, Caltech256, COCO, and LVIS datasets. While incrementing the value of $\ell$, we add embedders sorted by their weight in descending order (Higher performing embedders are being used first). Figure~\ref{fig:ablation-m-l} shows the mean average precision (MAP) achieved for each combination. In general, increasing both $m$ and $\ell$ leads to improved MAP. Notably, even the lowest configuration ($m=1$, $\ell=1$) outperforms most of the baselines evaluated in Table~\ref{tab:od-results} across different benchmarks.

From this ablation study, we conclude that consensus among embedders can lead to higher reliability and improved performance in \system. Note that \system supports an arbitrary number of embedders, allowing users to increase this number until performance saturates. This experiment also demonstrates that increasing the number of generated images enhances performance; specifically, the performance gap between $m=1$ and $m=2$ is larger than that between $m=2$ and $m=3$. This indicates that \system can achieve good results with a small number of images and does not require extensive image generation to capture all relevant complexities.

\stitle{Foundation Models}
\system supports image generation using various open-source and proprietary foundation models. These models differ in the quality of generated images, realism, and their alignment with the input prompt. In this section, we investigate two primary aspects. First, we assess the performance of each individual foundation model when used on its own. Second, we examine how combining multiple foundation models can improve performance by providing diverse representations of the input query.

Our experiments involve four foundation models: {\sc DALL-E3} (proprietary), {\sc Imagen3-fast} (proprietary), {\sc Flux-Schnell} (open-source), and {\sc RealViSXL} (based on Stable-diffusionXL; open-source). We evaluate their performance on object detection tasks using the BDD, Caltech256, and COCO datasets, as well as on complex natural language queries using the Winoground and COLA benchmarks. The inclusion of tests on complex natural language queries is motivated by the higher demands for prompt alignment and image quality in such scenarios. We calculate MAP for Object detection and Accuracy for Complex natural language benchmarks as the performance metric. Figure~\ref{fig:fms-performance} demonstrates how each foundation model performed in different benchmarks. 

Notably, the two open-source models, Flux-Schnell and RealVisXL, outperformed proprietary models by a huge margin in Complex queries, indicating they have better prompt alignment and understanding compared to DALL-E3 and Imagen3-fast. DALL-E3 performed worse than others in most of the benchmarks, and RealVisXL demonstrated to be a reliable option for most cases. 

Figure~\ref{fig:ablation-fms} illustrates that increasing the number of foundation models leads to an improvement in the overall performance of \system. Notably, the addition of further foundation models yields more improvement in performance for complex natural language queries than for object detection benchmarks. This outcome is likely attributable to the increased likelihood of generating images that accurately capture the nuanced details of the input prompt, thereby facilitating the correct identification of the target image. In contrast, for object detection tasks—where the objects are generally simpler to generate—the primary benefit of including additional foundation models is an increase in recall due to the diverse representations of the target object across various formats and styles.

\begin{figure}[!tb]
    \centering
    \footnotesize
    \begin{tikzpicture}
        \begin{groupplot}[
            group style={
                group size=2 by 1,         %
                horizontal sep=1.5cm,      %
            },
            width=0.5\columnwidth,      %
            xbar,
            enlarge y limits=0.15,
            nodes near coords,
            ylabel near ticks,
            xlabel near ticks,
            tick align=outside,
            yticklabel style={
              xshift=-0.3cm,
              align=center
            },
            legend style={
                at={(1,1.2)}, %
                anchor=south,
                legend columns=4,
                draw=none,
                font=\scriptsize
            },
        ]
        \nextgroupplot[
            title={\bf Object Detection Benchmarks}, 
            symbolic y coords={Caltech256,COCO,BDD},
            ytick=data,
            y dir=reverse,
            bar width=5pt,
            xlabel={MAP},
            enlarge y limits=0.35,
            enlarge x limits=0.2,
            y tick label style={rotate=90, anchor=center},
        ]

            \addplot+[xbar,fill=myorange!50, draw=myorange]
                coordinates {(0.689,BDD) (0.914,Caltech256) (0.959,COCO)};
            \addplot+[xbar,fill=mygreen!50, draw=mygreen]
                coordinates {(0.732,BDD) (0.917,Caltech256) (0.951,COCO)};
            \addplot+[xbar,fill=mypurple!50, draw=mypurple]
                coordinates {(0.608,BDD) (0.908,Caltech256) (0.926,COCO)};
            \addplot+[xbar,fill=myblue!50, draw=myblue]
                coordinates {(0.679,BDD) (0.943,Caltech256) (0.957,COCO)};

            \addlegendentry{\sc Imagen3-fast}
            \addlegendentry{\sc Flux-schnell}
            \addlegendentry{\sc DALL-E}
            \addlegendentry{\sc RealVisXL}

        \nextgroupplot[
            title={\bf Complex NLQ Benchmarks},
            symbolic y coords={COLA,Winoground},
            ytick=data,
            bar width=5pt,
            xlabel={Accuracy},
            y dir=reverse,
            y tick label style={rotate=90, anchor=center},
            enlarge y limits=0.5,
            enlarge x limits=0.2
        ]

            \addplot+[xbar,fill=myorange!50, draw=myorange]
                coordinates {(0.598,COLA) (0.434,Winoground)};
            \addplot+[xbar,fill=mygreen!50, draw=mygreen]
                coordinates {(0.623,COLA) (0.513,Winoground)};
            \addplot+[xbar,fill=mypurple!50, draw=mypurple]
                coordinates {(0.606,COLA) (0.435,Winoground)};
            \addplot+[xbar,fill=myblue!50, draw=myblue]
                coordinates {(0.614,COLA) (0.534,Winoground)};
        \end{groupplot}
    \end{tikzpicture}
    \caption{Comparison of Performance of Different Image Generators on Object Detection and Complex Natural Language Query benchmarks.}
    \label{fig:fms-performance}
\end{figure}

\begin{figure}[!tb]
    \centering
    \begin{subfigure}[b]{0.48\columnwidth}
        \centering
        \begin{tikzpicture}
            \begin{axis}[
                title={\bf Object Detection Benchmarks},
                width=\linewidth, %
                height=5cm,
                xlabel={\#Foundation Models},
                ylabel={MAP},
                xmin=0.5, xmax=4.5,
                ymin=0.65, ymax=1,
                xtick={1,2,3,4},
                legend style={
                    at={(0.5,1.30)}, %
                    anchor=south,
                    legend columns=4,
                    draw=none,
                    font=\small,
                },
                ymajorgrids=true,
                grid style=dashed
            ]

            \addplot+[mark=o,thick,color=mypurple]
                coordinates {
                    (1,0.690)
                    (2,0.709)
                    (3,0.703)
                    (4,0.720)
                };
            \addlegendentry{BDD}

            \addplot+[mark=x,thick,color=mygreen]
                coordinates {
                    (1,0.959)
                    (2,0.970)
                    (3,0.974)
                    (4,0.976)
                };
            \addlegendentry{COCO}

            \addplot+[mark=square,thick,color=myblue]
                coordinates {
                    (1,0.916)
                    (2,0.942)
                    (3,0.958)
                    (4,0.966)
                };
            \addlegendentry{Caltech256}

            \end{axis}
        \end{tikzpicture}
        \label{fig:plot1}
    \end{subfigure}
    \hfill
    \begin{subfigure}[b]{0.48\columnwidth}
        \centering
        \begin{tikzpicture}
            \begin{axis}[
                title={\bf Complex NLQ Benchmarks},
                width=\linewidth, %
                height=5cm,
                xlabel={\#Foundation Models},
                ylabel={Accuracy},
                xmin=0.5, xmax=4.5,
                ymin=0.5, ymax=0.7,
                xtick={1,2,3,4},
                legend style={
                    at={(0.5,1.3)}, %
                    anchor=south,
                    legend columns=2,
                    draw=none,
                    font=\small, %
                },
                ymajorgrids=true,
                grid style=dashed,
            ]

            \addplot+[mark=triangle,thick,color=myorange]
                coordinates {
                    (1,0.519)
                    (2,0.554)
                    (3,0.569)
                    (4,0.593)
                };
            \addlegendentry{Winoground}

            \addplot+[mark=diamond,thick,color=mylightgreen]
                coordinates {
                    (1,0.603)
                    (2,0.612)
                    (3,0.629)
                    (4,0.631)
                };
            \addlegendentry{COLA}

            \end{axis}
        \end{tikzpicture}
        \label{fig:plot2}
    \end{subfigure}
    \caption{The effect of number of Foundation Models on the Performance of \system for different Benchmarks}
    \label{fig:ablation-fms}
\end{figure}

\begin{figure}[!tb]
    \centering
    \footnotesize
    \begin{tikzpicture}
        \begin{groupplot}[
            group style={
                group size=2 by 1,
                horizontal sep=15mm,
                group name=myplots
            },
            width=0.5\columnwidth, %
            enlarge y limits=0.15,
            ylabel near ticks,
            xlabel near ticks,
            tick align=outside,
            legend style={
                at={(1,1.3)}, %
                anchor=south,
                legend columns=3,
                column sep=0.25cm,
                draw=none,
                font=\scriptsize
            },
            ymajorgrids=true,
            grid style=dashed,
        ]

        \nextgroupplot[
            title={\bf BDD},
            height=5cm, %
            xlabel={Image Size},
            ylabel={MAP}, %
            xmin=0.5, xmax=3.5,
            ymin=0.5, ymax=0.8,
            xtick={1,2,3},
            xticklabels={SMALL,MEDIUM,LARGE},
        ]
            \addplot[
                color=mypurple,
                mark=square,
                thick,
            ] coordinates {
                (1, 0.57)
                (2, 0.61)
                (3, 0.63)
            };
            \addplot[
                color=mygreen,
                mark=diamond,
                thick,
            ] coordinates {
                (1, 0.72)
                (2, 0.73)
                (3, 0.73)
            };
            \addplot[
                color=myblue,
                mark=o,
                thick,
            ] coordinates {
                (1, 0.64)
                (2, 0.68)
                (3, 0.69)
            };

            \legend{DALL-E, Flux-Schnell, RealVisXL}

        \nextgroupplot[
            title={\bf COCO},
            height=5cm, %
            xlabel={Image Size},
            xmin=0.5, xmax=3.5,
            ymin=0.8, ymax=1.0,
            xtick={1,2,3},
            xticklabels={SMALL,MEDIUM,LARGE},
        ]
            \addplot[
                color=mypurple,
                mark=square,
                thick,
            ] coordinates {
                (1, 0.881)
                (2, 0.931)
                (3, 0.935)
            };
            \addplot[
                color=mygreen,
                mark=diamond,
                thick,
            ] coordinates {
                (1, 0.941)
                (2, 0.952)
                (3, 0.955)
            };
            \addplot[
                color=myblue,
                mark=o,
                line width=1.5pt, %
            ] coordinates {
                (1, 0.932)
                (2, 0.961)
                (3, 0.964)
            };
        \end{groupplot}
    \end{tikzpicture}
    \caption{Illustration of variation of \system MAP based on size of guide image for different foundation models}
    \label{fig:ablation-imagesize}
\end{figure}

\stitle{Image Quality}
Foundation models usually support generating images in different resolutions. In this ablation study, we investigate the effect of image resolution (size) of guide images on \system's performance. We define three supported image sizes for each foundation model, \texttt{SMALL}, \texttt{MEDIUM} and \texttt{LARGE}. The exact amount of pixels is determined based on the foundation model design. We generate 3 images per query and utilize Flux-Schnell, RealVisXL and DALL-E for this study (ImagenV3-fast does not support different images sizes as input).
Figure~\ref{fig:ablation-imagesize} illustrates the MAP performance of various foundation models on the LVIS and COCO benchmarks across different image quality settings. It is evident that foundation models are typically fine-tuned for a specific image resolution, which in turn yields the best quality and prompt alignment. In our experiments, DALL-E demonstrated a consistent improvement in performance with increasing image sizes, whereas Flux-Schnell and RealVisXL exhibited a less pronounced correlation between performance and image size, particularly between the \texttt{MEDIUM} and \texttt{LARGE} settings.

\begin{figure}[!tb]
    \centering
    \small
    \begin{tikzpicture}
    \begin{axis}[
                width=0.5\textwidth,
                height=5cm,
				ybar,
				ylabel={Percentage (\%)},
                legend style={anchor=north east, draw=none, fill=none, font=\large},
				bar width=9pt,
				symbolic x coords={Needle,CLIP,Both,Neither},
				xtick=data,
                ylabel near ticks,
                xlabel near ticks,
                ymin=0, 
                ymax=70,  
                nodes near coords 
			]

    \addplot[fill=airforceblue, draw=blue] table [x=Engine, y=percentage, col sep=comma]
    {case_study.csv};
    \end{axis}
\end{tikzpicture}
    \caption{Results from the human evaluation case study on the \lvis dataset. Among the queries with a clear preference, \system was chosen in approximately 70\% of cases, underscoring its practical effectiveness in retrieving relevant images.}    
    \label{fig:case_study}
\end{figure}

\stitle{Query Complexity Classifier}\label{sec:eval_complexity_classifier}
This section details the empirical investigation into the impact of our proposed Query Complexity Classifier on the overall efficiency and retrieval performance of \system. The primary objective is to demonstrate how dynamically routing queries based on their predicted complexity can significantly reduce computational overhead without compromising retrieval effectiveness.

\begin{figure*}
    \centering
    \begin{subfigure}[b]{0.49\linewidth}
        \centering
        \includegraphics[height=0.25\textheight]{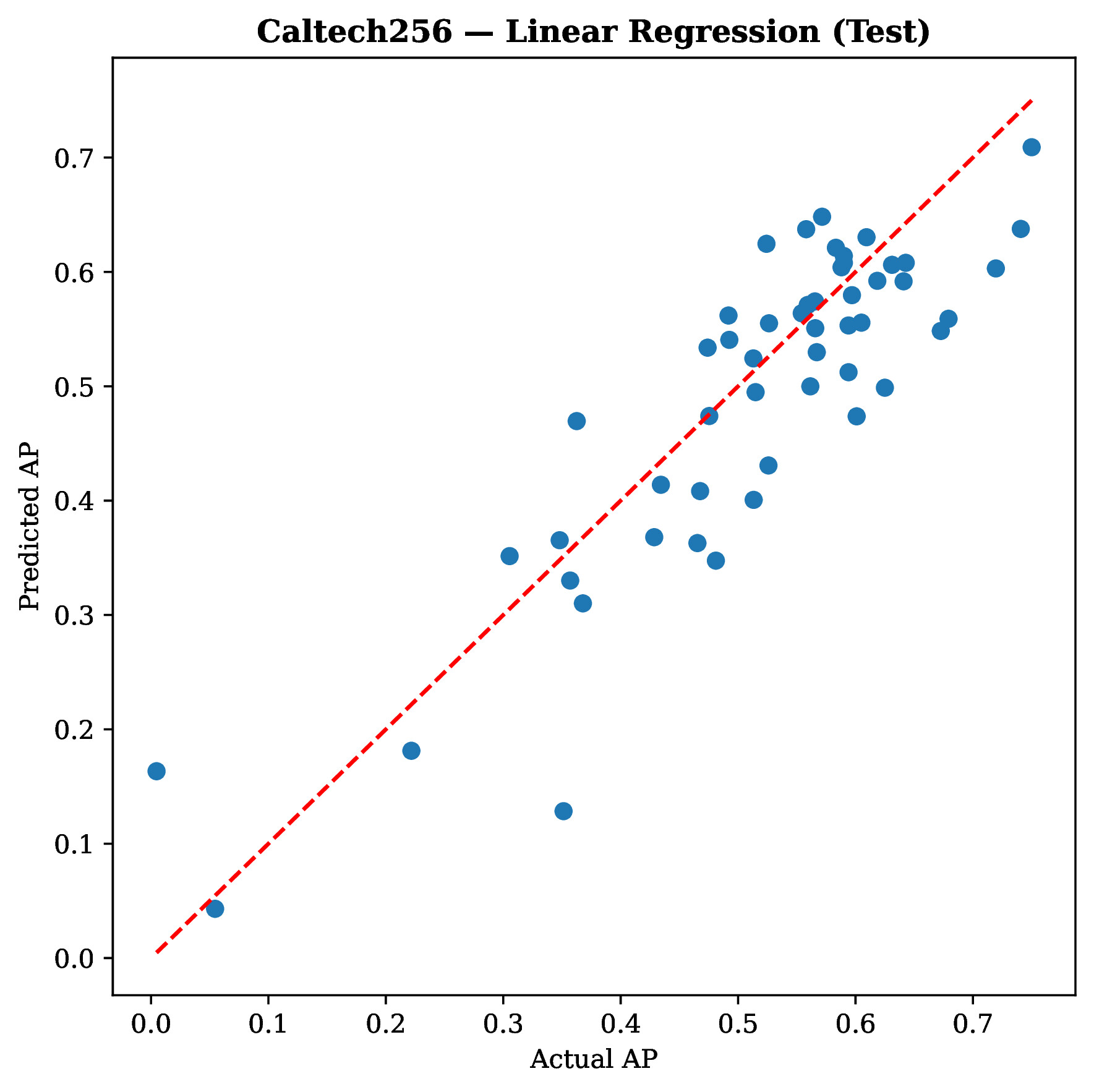}
        \caption{\small Comparison of Actual AP and Predicted AP using \system Query Complexity Classifier}
        \label{fig:qca_comparison}
    \end{subfigure}
    \hfill
    \begin{subfigure}[b]{0.49\linewidth}
        \centering
        \includegraphics[height=0.25\textheight]{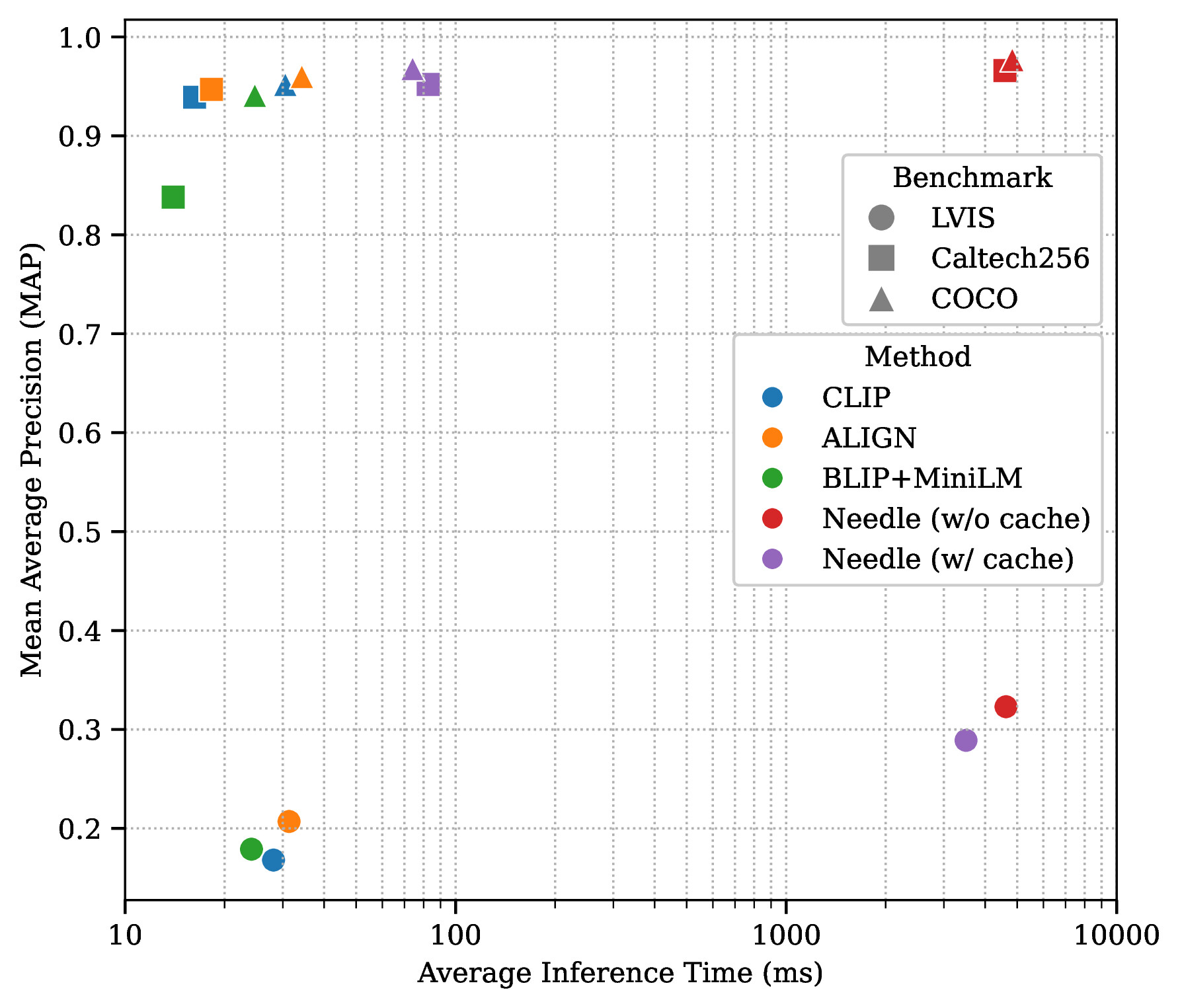}
        \caption{\small Comparison of Inference Speed and MAP using different baselines and benchmarks}
        \label{fig:inference_speed_comparison}
    \end{subfigure}
    \caption{Illustration of how \system Query Complexity Classifier improves \system inference speed}
    \label{fig:qca}
\end{figure*}

For these experiments, the query complexity classifier is implemented as a Linear Regression model, trained to predict the expected Average Precision (AP) of a given query when processed by preliminary retrieval methods. We specifically leverage features derived from two highly efficient preliminary methods: CLIP~\cite{radford2021learning} and ALIGN~\cite{jia2021scaling}. The features utilized for this prediction, as detailed in Section~\ref{sec:prac:complexity}, include mean top-$K$ cosine similarity scores, top-$K$ inter-method overlap coefficients, and confidence deviation for both CLIP and ALIGN.

The model was trained on a combined dataset derived from the training splits of the Caltech256~\cite{griffin_holub_perona_2022} and LVIS~\cite{gupta2019lvis} benchmarks. The ground-truth target for training was the actual AP achieved by the preliminary methods for each query. Figure~\ref{fig:qca_comparison} showcases the performance of our trained classifier in predicting AP for Caltech256. 

The primary benefit of the query complexity classifier is its ability to dynamically route queries, thereby enhancing the inference speed of \system. To quantify this improvement, we conducted experiments comparing the average inference speed and overall retrieval accuracy (Mean Average Precision, MAP) of \system, both with and without the classifier, against several established baselines. The baselines include: CLIP, ALIGN, and BLIP+MiniLM.
Our evaluation was performed across diverse benchmarks, including Caltech256, LVIS, and BDD100k~\cite{yu2020bdd100k}. Figure~\ref{fig:inference_speed_comparison} illustrates a comprehensive comparison of average inference speed and retrieval accuracy for all evaluated baselines and configurations of \system.

As one can observe from the results, employing our proposed query complexity classifier enables \system to achieve significantly faster inference speeds with only a minimal or negligible loss in overall retrieval performance. This is primarily attributed to the classifier's ability to accurately identify "simple" queries, for which the computationally intensive guide image generation process can be entirely bypassed. Furthermore, it is important to note that \system's efficiency is designed to progressively increase over time through its caching mechanism. As textual metadata for different images are extracted from user's queries (\ref{sec:prac:metadata}), the percentage of cache hits will naturally rise, further reducing the overhead of on-the-fly image generation and consequently enhancing inference speed.

\subsection{Case Study}\label{sec:exp:study}
To further validate \system's practical applicability, we conducted a case study with human evaluators. We utilized the \lvis dataset as our primary image repository, as it contains over 100k images spanning various categories, ensuring that any randomized query would likely have related images. This made \lvis an ideal choice for our case study.

Each human evaluator was asked to write five queries of their choice, with varying levels of complexity. For each query, we retrieved 20 relevant images from both \clip and \system, displaying the results side by side with randomized orientations for each query. Evaluators were then asked to select the more relevant results, choosing from four options: "Left is better," "Right is better," "Both are good enough," and "Neither is good enough". Then, we recorded the engine behind that option as the preferred engine. 

A total of 20 human evaluators participated in the study, contributing 100 queries. The queries ranged from simple ones like "Celery stew" to more complex descriptions such as "A person walking on the sidewalk of a river in Fall at Chicago.", Figure~\ref{fig:case_study} presents the results of this user study. If we consider only the queries where one engine is preferred over the other, \system is chosen as the better engine in almost 70\% of cases.

\end{document}